\documentclass[a4paper,11pt]{article}
\usepackage{pos}

\title{New results in AdS/CFT in low dimensions from massive Type IIA}
\ShortTitle{AdS$_3$/CFT$_2$ and AdS$_2$/CFT$_1$}

\author*[a,b]{Yolanda Lozano}
\author[c]{Nicolò Petri}
\author[a,b,d]{Anayeli Ramirez}

\affiliation[a]{Department of Physics, University of Oviedo,\\
Avda. Federico Garcia Lorca s/n, 33007 Oviedo, Spain}

\affiliation[b]{Instituto Universitario de Ciencias y Tecnolog\'ias Espaciales de Asturias (ICTEA),\\
Calle de la Independencia 13, 33004 Oviedo, Spain}

\affiliation[c]{Department of Physics, Ben-Gurion University of the Negev,\\ Beer Sheva 84105, Israel}

\affiliation[d]{Kavli Institute for Theoretical Physics, \\University of California, Santa Barbara, CA 93106}

\emailAdd{ylozano@uniovi.es}
\emailAdd{petri@post.bgu.ac.il}
\emailAdd{ramirezanayeli.uo@uniovi.es}

\abstract{We review recent developments in the study of the AdS/CFT correspondence in low dimensions, focusing on the construction of AdS$_3$/CFT$_2$ and AdS$_2$/CFT$_1$ dual pairs in massive Type IIA string theory. We start discussing the $\text{AdS}_3\times S^2\times \text{CY}_2\times I$ solutions to massive IIA supergravity with $\mathcal{N}=(0,4)$ supersymmetry constructed in \cite{Lozano:2019emq}. We review the 2d CFTs dual to these solutions, together with their defect interpretation as surface defects within 5d fixed point theories living in D4-D8 bound states. Next, we discuss the $\text{AdS}_2\times S^3\times \text{CY}_2\times I$ solutions with $\mathcal{N}=4$ supersymmetry constructed in \cite{Lozano:2020sae}. We discuss the superconformal quantum mechanics dual to these solutions, that we interpret in terms of line defects realised as D0-D4 baryon vertices in 5d Sp(N) fixed point theories. We review a particular example in this class of solutions, constructed through non-Abelian T-duality with respect to a non-compact isometry group, and discuss the possible embedding of a 3d black hole geometry constructed long ago via non-Abelian T-duality within this solution.}

\FullConference{%
  Corfu Summer Institute 2021 "School and Workshops on Elementary Particle Physics and Gravity"\\
  29 August - 9 October 2021\\
  Corfu, Greece
}

\begin{document}
\maketitle

\section{Introduction}
The study of AdS$_3$ and AdS$_2$ spaces in String Theory has been of paramount importance towards achieving our current microscopical understanding of black holes.  These spaces describe the geometries of extremal black holes close to the horizon, and through the AdS/CFT correspondence have associated dual CFTs where their microscopical degrees of freedom can be identified. The agreement between the field theory degrees of freedom and the Bekenstein-Hawking entropy represents one of the most important achievements of String Theory in the last decades. 

Recently, remarkable progress has been gained in the construction of AdS$_3$ and AdS$_2$ solutions to Type II supergravities for which the dual CFTs have also been identified \cite{Lozano:2019emq}-\cite{Couzens:2021veb}. These represent explicit new AdS/CFT pairs where the black hole microscopical counting program can be carried out in detail. In the AdS$_2$/CFT$_1$ case the well-known problems related to the non-connectedness of the boundary of AdS$_2$ and the interpretation of the central charge of the dual super-conformal quantum mechanics (SCQM) have been circumvented through explicit constructions of SCQM whose degrees of freedom match the Bekenstein-Hawking entropy \cite{Lozano:2020txg,Lozano:2020sae,Lozano:2021rmk,Ramirez:2021tkd,Lozano:2021fkk}.

Low dimensional AdS spaces constitute as well promising candidates to holographic duals of CFTs describing defects within higher dimensional CFTs. Notable examples of such realisations have been reported in \cite{Argurio:2000tg,Karch:2000gx,DeWolfe:2001pq,Aharony:2003qf,DHoker:2006vfr,Lunin:2007ab,DHoker:2007zhm,DHoker:2007mci,Chiodaroli:2009yw,Chiodaroli:2009xh,Dibitetto:2017tve,Dibitetto:2017klx,Dibitetto:2018iar,Dibitetto:2018gtk,Chen:2020mtv,Faedo:2020nol,Faedo:2020lyw,Gutperle:2020rty,Lozano:2021fkk}. In these realisations the brane set-ups in which the defect CFTs live are interpreted as brane intersections ending on bound states, which are described close to the horizon by higher dimensional AdS spaces. The brane intersections break some of the isometries of these higher dimensional  spaces, giving rise to lower dimensional AdS spaces in the near horizon limit. These lower dimensional spaces are dual to low dimensional CFTs, that find an interpretation as defect CFTs within the higher dimensional CFTs living in the bound states on which the brane intersections end.

In these proceedings we will report on recent progress in the construction of AdS$_3$/CFT$_2$ and AdS$_2$/CFT$_1$ pairs in massive Type IIA supergravity while paying special attention to the description of the CFTs and their defect interpretation. The AdS solutions are foliations of $\text{AdS}_3\times S^2$ or $\text{AdS}_2\times S^3$ times a $\text{CY}_2$ over an interval, preserving 4 supersymmetries\footnote{More concretely, $\mathcal{N}=(0,4)$ in the  AdS$_3$/CFT$_2$ case.}. Remarkably, the CFTs dual to general subclasses of these solutions have been shown to admit quiver descriptions in the UV which have been used to compute their degrees of freedom, which have been shown to match the holographic computations. These solutions thus constitute well-defined string theory settings where computations such as the corrections to the entropy of five and four dimensional black holes can be performed.

The approach taken in the construction of the AdS$_2$ solutions is to apply double analytical continuation techniques on the AdS$_3$ solutions. Compared to other approaches in the literature (see for instance \cite{Lozano:2020txg,Lozano:2021rmk,Lozano:2021fkk}) this allows to construct AdS$_2$ spaces unrelated to AdS$_3$ ones, and therefore dual to SCQMs that do not occur as discrete light-cone compactifications of 2d CFTs \cite{Balasubramanian:2003kq,Balasubramanian:2009bg}. Yet, the explicit dual pairs that we will review show that it is possible to compute the SCQM central charge from a formula inherited from 2d.
This is a striking result that deserves more detailed investigation. 

In this article we will also put our focus on the interpretation of the new dual pairs as describing defect CFTs within the 5d Sp(N) fixed point theories living in D4-D8 bound states \cite{Seiberg:1996bd}, whose near horizon geometry is the   Brandhuber-Oz AdS$_6$ solution to massive Type IIA supergravity \cite{Brandhuber:1999np}. The approach that we take is to relate (a subset of) our solutions with the uplift to massive IIA of the AdS$_3$ and AdS$_2$ domain wall solutions to 6d minimal gauged supergravity obtained in \cite{Dibitetto:2017klx,Dibitetto:2018iar,Dibitetto:2018gtk}. These solutions asymptote locally to the AdS$_6$ vacuum in the UV, while they are singular in the IR, due to the presence of lower dimensional brane intersections. Thus, they can be interpreted as holographic duals of surface (for AdS$_3$) or line (for AdS$_2$) defect CFTs within the 5d Sp(N) fixed point theory dual to the AdS$_6$ vacuum.

The paper is organised as follows. We start in section \ref{seed} by reviewing the $\text{AdS}_3\times S^2\times \text{CY}_2\times I$ solutions constructed in \cite{Lozano:2019emq}, with a focus on the subclass for which the dual 2d CFT was identified in \cite{Lozano:2019jza,Lozano:2019zvg,Lozano:2019ywa}\footnote{Here we follow closely \cite{Couzens:2021veb}, where some errors in the field theory description in \cite{Lozano:2019jza,Lozano:2019zvg,Lozano:2019ywa} were pointed out and a more careful analysis of the matching between the field theory and holographic central charges was carried out.}. Then in subsection \ref{defectAdS3} we describe the defect interpretation of these solutions within the 5d Sp(N) fixed point theory, found in \cite{Faedo:2020nol}. In section  \ref{TypeC} we turn to the study of the $\text{AdS}_2\times S^3\times \text{CY}_2\times I$ solutions constructed in \cite{Lozano:2020bxo,Lozano:2020sae}, with special focus on the subclass of solutions for which the dual SCQM was identified. We devote subsection \ref{NATD} to review a solution in the AdS$_2$ class recently constructed in \cite{Ramirez:2021tkd}, by means of a non-Abelian T-duality (NATD) transformation acting on the $\text{AdS}_3\times S^3\times \text{CY}_2$ solution of Type IIB string theory. Contrary to previous applications of NATD as a solution generating technique in supergravity, the NATD takes place in this case with respect to a non-compact group of isometries, mapping the $\text{AdS}_3\times S^3\times \text{CY}_2$ space onto an $\text{AdS}_2\times I'\times S^3\times \text{CY}_2$ solution contained in the class of \cite{Lozano:2020bxo,Lozano:2020sae}. We try to connect the previous solution to a black hole geometry constructed in \cite{Alvarez:1993qi}, when non-Abelian T-duality was first introduced at the level of the string worldsheet. This geometry was found by performing NATD on the principal chiral model with group SL(2,$\mathbf{R}$), as an illustration of the applicability of NATD with respect to non-compact isometry groups. Our results in this subsection show that this black hole geometry cannot be embedded within massive Type IIA supergravity using the class of solutions constructed in \cite{Lozano:2020bxo,Lozano:2020sae}. These are new results in our search for valid string theory backgrounds where the black hole geometry constructed in \cite{Alvarez:1993qi} could be embedded. In subsection \ref{defectAdS2} we turn to the defect interpretation of (a subclass of) the $\text{AdS}_2\times S^3\times \text{CY}_2\times I$ solutions as line defects within the 5d Sp(N) fixed point theory, following \cite{Faedo:2020nol}. Finally in section \ref{discu} we summarise the contents of this paper and sketch future new directions of investigation.

 \section{AdS$_3$/CFT$_2$ with (0,4) supersymmetries}\label{seed}
 
 In \cite{Lozano:2019emq} a family  of AdS$_3\times$S$^2$ solutions to massive IIA supergravity with $\mathcal{N}=(0,4)$ supersymmetry and SU(2)-structure was constructed. These solutions are foliations of AdS$_3\times$S$^2\times$M$_4$ over an interval, where M$_4$ is either a CY$_2$ or a 4d K\"{a}hler manifold. Both cases were studied in detail in \cite{Lozano:2019emq}. In this review article we will focus on the case  M$_4=$CY$_2$, referred as Class I in that reference.  
 
The Neveu-Schwarz sector of this subclass of solutions reads, in string frame\footnote{Note that we are restricting to the case in which the closed and anti-self dual 2-form living on the $\text{CY}_2$ also included in \cite{Lozano:2019emq} vanishes.},
\begin{equation}
\begin{split}
	\mathrm{d}s^2&= \frac{u}{\sqrt{h_4 h_8}}\bigg(\mathrm{d}s^2_{\text{AdS}_3}+\frac{h_8h_4}{\Delta} \mathrm{d}s^2_{\text{S}^2}\bigg)+ \sqrt{\frac{h_4}{h_8}} \mathrm{d} s^2_{\text{CY}_2}+ \frac{\sqrt{h_4 h_8}}{u} \text{d}z^2,\qquad \Delta=4 h_8h_4+u'^2\;,\label{seedNS}\\
	e^{-\Phi}&= \frac{h_8^{\frac{3}{4}} }{2h_4^{\frac{1}{4}}\sqrt{u}}\sqrt{\Delta}\;,\qquad~~~~ H_3= \frac{1}{2} \mathrm{d} \left(-z+\frac{ u u'}{\Delta}\right)\wedge\text{vol}_{\text{S}^2},
\end{split}
\end{equation}
where $\Phi$ is the dilaton and  $H_3
$ is the field strength of the Kalb-Ramond antisymmetric tensor, $B_2
$. The warping functions $h_8$ and $u$ have support on the $z$ coordinate while $h_4$ has support on $(z, \text{CY}_2)$. We have denoted $u'= \partial_{z}u$, and the same for the functions $h_4$ and $h_8$ below. The background \eqref{seedNS} is supported by the Ramond-Ramond (RR) fluxes,
\begin{equation}
	\begin{split}
		F_0&=h_8',\;\;\;\qquad F_2=-\frac{1}{2}\bigg(h_8- \frac{ h'_8 u'u}{\Delta} \bigg)\text{vol}_{\text{S}^2},\\[2mm]
		F_4&= -\bigg(\mathrm{d}\left(\frac{u u'}{2h_4}\right)+2 h_8  \text{d}z\bigg) \wedge\text{vol}_{\text{AdS}_3}- \partial_z h_4\text{vol}_{\text{CY}_2}-h_8 (*_4d_4h_4)\wedge dz.
		\label{RRsector-seed}
	\end{split}
\end{equation}
Additionally, supersymmetry demands  
\begin{equation} 
u''(z)=0,\label{eqsmotion1}
\end{equation}
and away from localised sources, the Bianchi identities demand,  
\begin{equation} 
	h_8''(z)=0,\;\;\;\; \partial_z^2 h_4+h_8\nabla^2_{\text{CY}_2}h_4=0.\label{eqsmotion}
\end{equation}
It was shown in \cite{Lozano:2019emq} that the background defined by \eqref{seedNS}-\eqref{RRsector-seed} is a solution  of massive IIA supergravity preserving $(0,4)$ supersymmetries as long as the $h_4,h_8,u$ functions satisfy the conditions \eqref{eqsmotion1}-\eqref{eqsmotion}.

The  Page fluxes, defined as  $\hat{F}=e^{-B_2}\wedge F$,  are given by,
\begin{gather}
	\begin{split}
	\hat{F}_0&=h_8',\qquad\qquad
	\hat{F}_2=-\frac{1}{2}\bigg(h_8- h_8'(z-2\pi k)\bigg)\text{vol}_{\text{S}^2},\\
	\hat{F}_4&=-\bigg(\partial_z\left(\frac{u u'}{2 h_4}\right)+2 h_8\bigg)  \text{d}z \wedge\text{vol}_{\text{AdS}_3}
	-  \partial_z h_4\text{vol}_{\text{CY}_2}-h_8 (*_4d_4h_4)\wedge dz.\label{eq:background}
	\end{split}
\end{gather}
Here we have taken into account large gauge transformations of $B_2$ of parameter $k$, $B_2\to B_2 + { \pi k} \text{vol}_{\text{S}^2}$, for $k=0,1,...., P$, that ensure that it remains in the fundamental region,\begin{equation}
\frac{1}{4\pi^2}|\int_{\text{S}^2}B_2|\in [0,1].
\end{equation}
These transformations are performed every time a $z$-interval $z\in [2\pi k, 2\pi(k+1)]$ is crossed.

In the case in which $h_4$ does not depend on the coordinates of the $\text{CY}_2$, the conditions (\ref{eqsmotion}) leave us with linear functions for both $h_8$ and $h_4$. The analysis of the dual field theory carried out in \cite{Lozano:2019jza,Lozano:2019zvg,Lozano:2019ywa} considered functions of the form,
\begin{gather} \label{profileh4sp}
	h_4(z)\!=
	\;\!\!
	\left\{ \begin{array}{cccrcl}
		\frac{\beta_0 }{2\pi}
		z & 0\leq z\leq 2\pi, &\\
		\alpha_k\! +\! \frac{\beta_k}{2\pi}(z-2\pi k) &~~ 2\pi k\leq z \leq 2\pi(k+1),& ~~k=1,...,P-1\\
		\alpha_P-  \frac{\alpha_P}{2\pi}(z-2\pi P) &~~ 2\pi P\leq z \leq 2\pi(P+1),&
	\end{array}
	\right.\\
	\label{profileh8sp}
	h_8(z)
	=\left\{ \begin{array}{cccrcl}
		\frac{\nu_0 }{2\pi}
		z & 0\leq z\leq 2\pi, &\\
		\mu_k+ \frac{\nu_k}{2\pi}(z-2\pi k) &~~ 2\pi k\leq z \leq 2\pi(k+1),& ~~k=1,...,P-1\\
		\mu_P-  \frac{\mu_P}{2\pi}(z-2\pi P) &~~ 2\pi P\leq z \leq 2\pi(P+1),&
	\end{array}
	\right.
\end{gather}
which, being piecewise linear, allow for D4 and D8 sources in the background, as implied by the expressions for $\hat{F}_4$ and $\hat{F}_0$ in \eqref{eq:background}. Here it has been imposed that $h_4$ and $h_8$ vanish at $z=0$, where the space begins, and at $z=2\pi(P+1)$, where the space ends. The singularity structure of the metric and dilaton at  these points is that of a superposition of D2-branes wrapped on $\text{AdS}_3$ and smeared on the $\text{CY}_2\times \text{S}^2$, and D6-branes wrapped on $\text{AdS}_3\times \text{CY}_2$\footnote{In fact, it is also compatible with a superposition of O2-O6 planes. The string theory interpretation of smeared orientifold fixed planes is however unclear.}. 
In turn, $u$ needs to be continuous for preservation of supersymmetry. In this paper we will consider the simplest case $u'=0$. Those readers interested in the $u'\neq0$ case are referred to  \cite{Lozano:2019ywa, Dibitetto:2020bsh}.

Imposing the continuity of the Neveu-Schwarz sector across the various intervals one finds that the quantities $(\alpha_k,\beta_k, \mu_k, \nu_k)$ must satisfy,
 \begin{equation}
	\alpha_k=\sum_{j=0}^{k-1} \beta_j ,~~~\mu_k= \sum_{j=0}^{k-1}\nu_j.\label{definitionmupalphap}
\end{equation} 
In turn, the quantised charges are given, in the $[z_k,z_{k+1}]$ interval, by
\begin{eqnarray}
&&Q_{\text{D2}}^{(k)}=\alpha_k=\sum_{j=0}^{k-1}\beta_j, \qquad Q_{\text{D6}}^{(k)}=\mu_k=\sum_{j=0}^{k-1}\nu_j\nonumber\\
&&Q_{\text{D4}}^{(k)}=\beta_k,\qquad Q_{\text{D8}}^{(k)}=\nu_k,\qquad Q_{\text{NS5}}^{(k)}=1, \label{quantised-charges-1}
\end{eqnarray}
which implies that $(\alpha_k,\beta_k, \mu_k, \nu_k)$ must be integer numbers.
 
 In the next subsection we briefly summarise the two dimensional CFTs proposed in \cite{Lozano:2019jza,Lozano:2019zvg} as duals to the family of solutions given by \eqref{seedNS}-\eqref{RRsector-seed} with $h_4$, $h_8$ given by \eqref{profileh4sp}-\eqref{profileh8sp}.

 \subsection{Two dimensional dual CFTs}\label{CFT-seed}
 \begin{table}[t]
	\begin{center}
		\begin{tabular}{| l | c | c | c | c| c | c| c | c| c | c |}
			\hline		    
			& x$^0$ & x$^1$ & x$^2$ & x$^3$ & x$^4$ & x$^5$ & x$^6$ & x$^7$ & x$^8$ & x$^9$ \\ \hline
			D2 & x & x & &  &  &  & x  &   &   &   \\ \hline
			D4 & x & x &  &  &  &   &  & x & x & x  \\ \hline
			D6 & x & x & x & x & x & x & x  &   &   &   \\ \hline
			D8 & x & x &x  & x & x &  x &  & x & x & x  \\ \hline
			NS5 & x & x &x  & x & x & x  &   &   &  &  \\ \hline
		\end{tabular} 
	\end{center}
	\caption{Brane set-up underlying the background given by \eqref{seedNS}-\eqref{eqsmotion}. $(x^0,x^1)$ are the directions where the two dimmensional CFT lives. The directions $(x^2, \dots, x^5)$ span the CY$_2$, on which the D6- and the D8-branes are wrapped. The coordinate $x^6$ is the direction associated with $z$. Finally $(x^7,x^8,x^9)$ are the orthogonal directions realising the SO(3) R-symmetry.
	}   
	\label{D6-NS5-D8-D2-D4-first}	
\end{table} 

The branes that underlie the background
 defined by equations \eqref{seedNS}-\eqref{eqsmotion} are distributed as indicated in Table \ref{D6-NS5-D8-D2-D4-first}. The D2- and D6-branes play the role of colour branes, while the D4- and D8-branes are flavour branes.  This interpretation  is supported by the study of the Bianchi identities, given by 
 \begin{eqnarray}
&&\text{d}F_0=\sum_{k=1}^P\Bigl(\frac{\nu_{k-1}-\nu_k}{2\pi}\Bigr)\delta (z-2\pi k)\text{d}z\nonumber\\
&&\text{d}{\hat F}_4=\sum_{k=1}^P\Bigl(\frac{\beta_{k-1}-\beta_k}{2\pi}\Bigr)\delta (z-2\pi k)\text{d}z\wedge 
\text{vol}_{\text{CY}_2}\, , \label{Bianchis}
\end{eqnarray}
which show that at the points $z=2\pi k$ there are D4 and D8 localised sources. 

\begin{figure}[t!]
 	\centering
 	{{\includegraphics[width=10cm]{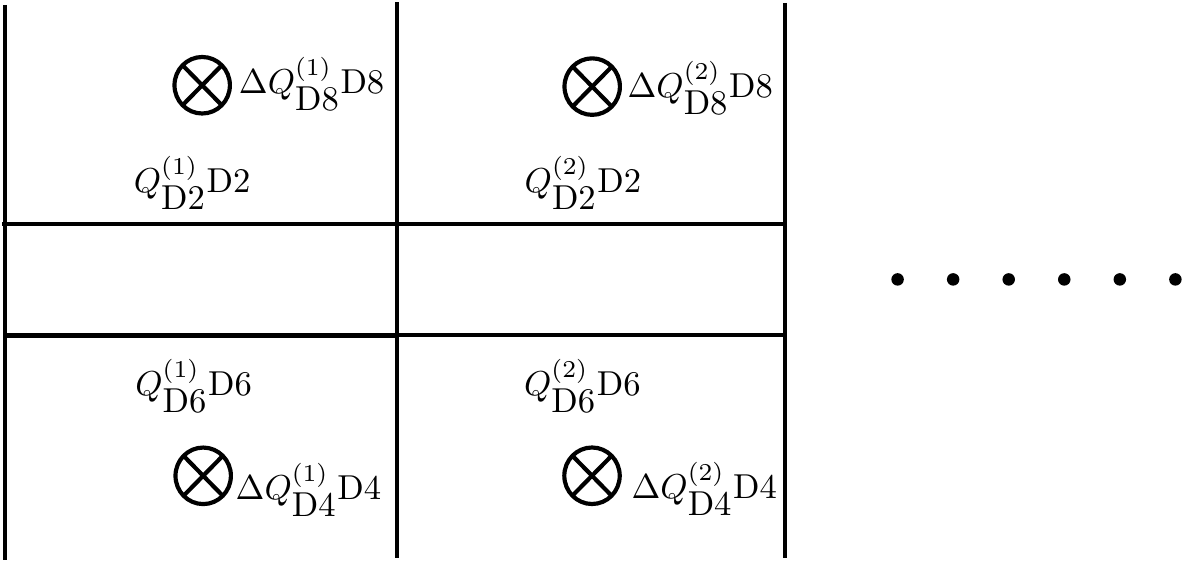} }}%
 	\caption{Hanany-Witten brane set-up associated to the solutions with $h_4$, $h_8$ functions given by \eqref{profileh4sp}-\eqref{profileh8sp}. The horizontal lines represent colour branes, in our case D2 and D6 branes, vertical lines represent NS5-branes and the crosses are flavour branes (D4- and D8-branes).  	
 	}	
	 	\label{HWbranesetup}
 \end{figure}
\begin{figure}[t!]
 	\centering
 	{{\includegraphics[width=10cm]{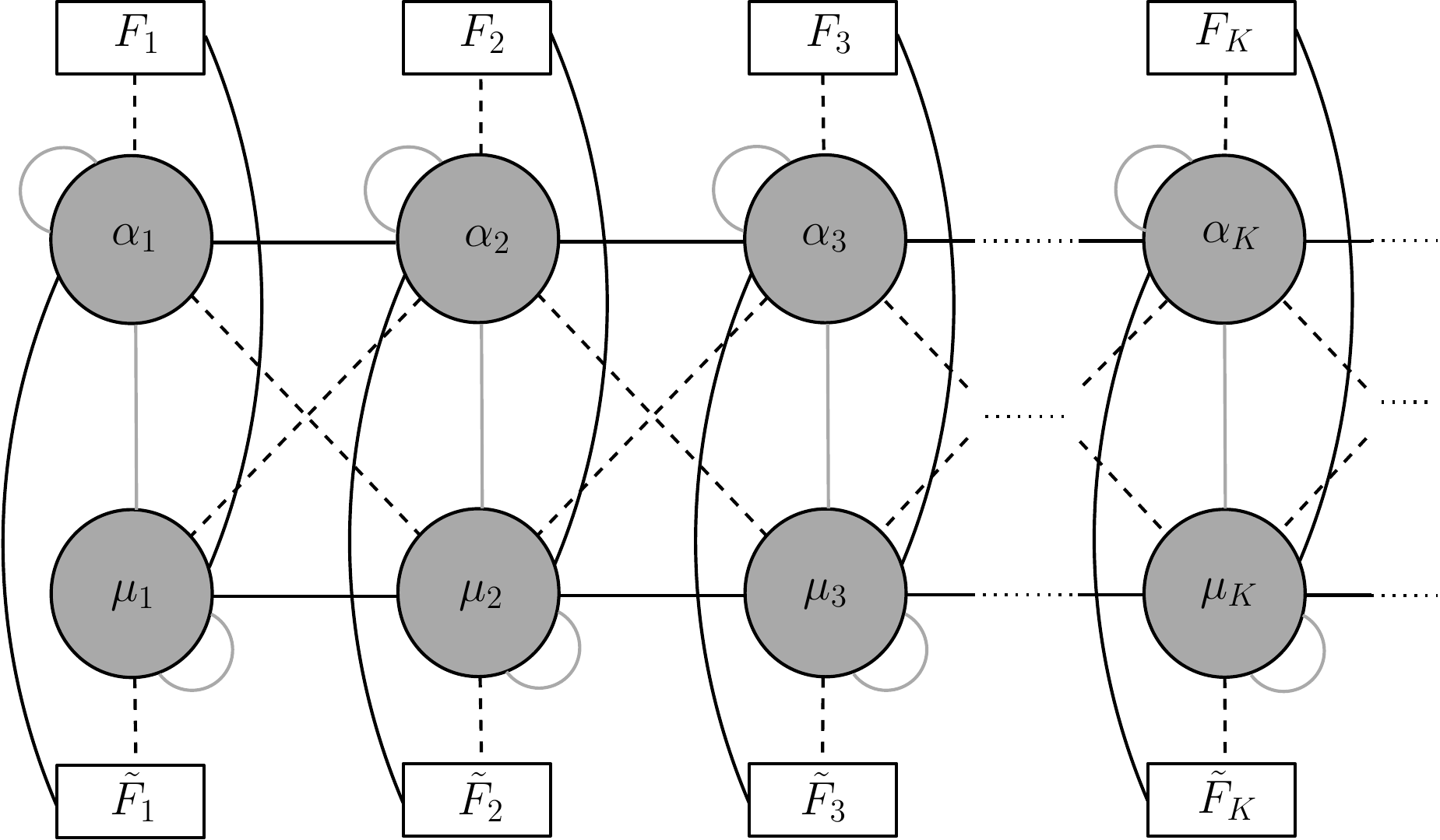} }}%

 	\caption{ Quivers encoding the two dimensional field theories living in the D2-D4-D6-D8-NS5 brane intersections depicted in Figure \ref{HWbranesetup}.
 	}
 	\label{figurageneral}
 \end{figure}

 The previous information can be codified in the Hanany-Witten brane set-up depicted in Figure \ref{HWbranesetup}. As shown in \cite{Lozano:2019jza,Lozano:2019zvg,Couzens:2021veb}, the 2d field theories living in these brane intersections are represented by the quivers depicted in Figure \ref{figurageneral}, whose dynamics conjecturally flow in the IR to CFTs with small ${\cal N}=(0,4)$ supersymmetry, dual to the AdS$_3$ solutions.
The 2d field theory lives in the D2 and D6 colour branes and there are adequate flavour groups coming from D4 and D8 branes, that give rise to non-anomalous quivers. 
 
 The quiver dynamics was first studied in \cite{Lozano:2019jza,Lozano:2019zvg}, and later analysed in deeper detail in \cite{Couzens:2021veb}, where the explicit quantisation of the open strings connecting the different branes in the set-up was carried out. This detailed analysis led to corrections to some of the results in \cite{Lozano:2019jza,Lozano:2019zvg}, not changing however significantly the main conclusions in these papers. The multiplets that arise from the quantisation of open strings are summarised in Table \ref{table:multiplets from strings}. The quivers depicted in Figure \ref{figurageneral} are then described in terms of $(0,4)$ vector multiplets and $(0,4)$ adjoint hypermultiplets, associated to the gauge nodes (depicted by circles and grey lines starting and ending on the same gauge group, respectively), $(4,4)$ twisted hypermultiplets in the bifundamental representation of two gauge groups (depicted by black lines), (0,4) bifundamental hypermultiplets (grey lines) and (0,2) bifundamental Fermi multiplets (dashed lines)
 . 

\begin{table}[h]
\begin{center}
\begin{tabular}{|c|c|c|c|}
\hline
String & Interval & Multiplet & Representation\\
\hline\hline
D2-D2 & Same & $\mathcal{N}=(0,4)$ vector $+$ $\mathcal{N}=(0,4)$ hyper & Adjoint\\
\hline
D2-D2 & Adjacent & $\mathcal{N}=(4,4)$ twisted hyper & bi-fundamental\\
\hline
D6-D6 & Same & $\mathcal{N}=(0,4)$ vector $+$ $\mathcal{N}=(0,4)$ hyper & Adjoint\\
\hline
D6-D6 & Adjacent & $\mathcal{N}=(4,4)$ twisted hyper & bi-fundamental\\
\hline
D2-D6 & Same &$\mathcal{N}=(0,4)$ hyper & bi-fundamental\\
\hline
D2-D6 & Adjacent & $\mathcal{N}=(0,2)$ Fermi & bi-fundamental\\
\hline
D2-D4 & Same & $\mathcal{N}=(4,4)$ twisted hyper & bi-fundamental\\
\hline
D4-D6 & Same & $\mathcal{N}=(0,2) $ Fermi & bi-fundamental\\
\hline
D2-D8 & Same & $\mathcal{N}=(0,2) $ Fermi &bi-fundamental \\
\hline
D6-D8 & Same & $\mathcal{N}=(4,4)$ twisted hyper & bi-fundamental\\
\hline
\end{tabular}
\end{center}
\caption{Summary of the multiplets arising from the different strings stretching between branes in the brane set-up. The interval column determines whether the branes lie in the same interval or in adjacent intervals.
For strings that do not contribute massless modes we have ignored their contribution in the table, for example D4-D4 strings.}
\label{table:multiplets from strings}
\end{table}

As discussed in  \cite{Lozano:2019jza,Lozano:2019zvg,Couzens:2021veb}, the cancellation of gauge anomalies constrains, for generic U($\alpha_k$) and U($\mu_k$) colour groups, the ranks of the respective flavour groups to be,
 \begin{equation}
 	F_k=\nu_{k-1}-\nu_{k},\;\;\;\; \tilde{F}_k=\beta_{k-1}-\beta_{k},\label{flavours}
 \end{equation}
 exactly as implied by \eqref{Bianchis}. Moreover, the field theory and holographic central charges can be shown to match in the holographic limit. Indeed, the right-moving central charge of the IR SCFT can be calculated using its relation with the U$(1)_R$ current two-point function,
 \begin{equation}
 c_R=3\text{Tr} [\gamma^3 Q_R^2],
 \end{equation}
 where the trace is over the Weyl fermions of the theory and $\gamma^3$ is the chirality matrix in 2d.
 Keeping in mind the R-charges and fermion content of the different multiplets, summarised in Table \ref{R-charges}, this leads to  
  \begin{equation}
 	c_R=6 (n_{hyp}- n_{vec}),\label{centralcft}
 \end{equation}
 where $n_{hyp}$ is the number of $(0,4)$ hypermultiplets and $n_{vec}$ is the number of $(0,4)$ vector multiplets.
  Note that (0,4) twisted hypermultiplets and (0,2) Fermi multiplets do not contribute to the R-symmetry anomaly, and therefore they do not contribute either to the central charge. For the quivers depicted in Figure \ref{figurageneral} the central
charge is then given by
\begin{equation}
c_R=6\sum_{k=1}^P \alpha_k\mu_k.
\end{equation}

 \begin{table}[h]
\begin{center}
\begin{tabular}{|c|c|c|c|c|}
\hline
Multiplet & $(0,2)$ Origin &  Number of Fermions & Chirality& R-charge of Fermion\\
\hline \hline
$(0,4)$ hyper& 2 $\times$ Chiral& 2 & R.H. & -1\\
\hline
$(0,4)$ twisted hyper& 2$\times$ Chiral & 2 & R.H. & 0\\
\hline
$(0,4)$ vector& (0,2) vector & 1 & L.H. & 1\\
&(0,2) Fermi &1 & L.H. & 1\\
\hline
$(0,2)$ Fermi & -  & 1& L.H. & 0\\
\hline
\end{tabular}
\end{center}
\caption{R-charges and fermion content of the multiplets.}
\label{R-charges}
\end{table}

In turn, the holographic central charge for the geometries defined by \eqref{seedNS} is given by
\begin{equation}
	c_{hol}= \frac{3\pi}{2 G_N}\text{Vol}_{\text{CY}_2}\int_0^{2\pi (P+1)} h_4 h_8 \text{d}z= \frac{3}{\pi} \int_0^{2\pi (P+1)} {h}_4 h_8 \text{d}z  .\label{chol}
\end{equation}
 Here we have used that $G_N= 8\pi^6$, with $g_s=\alpha'=1$, and that $ \text{Vol}_{\text{CY}_2}=16\pi^4$. For the functions $h_4$, $h_8$ displayed in \eqref{profileh4sp}-\eqref{profileh8sp} this gives
 \begin{equation} \label{holocc}
 c_{hol}=\sum_{k=1}^P\Bigl(6\alpha_k\mu_k+3(\alpha_k\nu_k+\beta_k\mu_k)+2\beta_k\nu_k\Bigr)\, .
 \end{equation}
As discussed in \cite{Couzens:2021veb}, this quantity has to be matched with the combination of left-moving and right-moving central charges of the field theory,
\begin{equation}
c_{hol}=\frac{c_L+c_R}{2}.
\end{equation}
The left-moving central charge can be computed from the field theory using that
\begin{equation}
c_L-c_R=\text{Tr}\gamma^3.
\end{equation}
This gives for the quivers depicted in Figure \ref{figurageneral},
\begin{equation}
c_L-c_R=\sum_{k=1}^P \Bigl( \alpha_k (\mu_{k+1}-\mu_k)+\mu_k(\alpha_{k+1}-\alpha_k)\Bigr),
\end{equation}
and finally
\begin{equation}
\frac{c_L+c_R}{2}=\sum_{k=1}^P\Bigl( 6\alpha_k\mu_k+\frac12(\alpha_k(\mu_{k+1}-\mu_k)+\mu_k(\alpha_{k+1}-\alpha_k))\Bigr).
\end{equation}
Comparing this expression to the expression \eqref{holocc} for the holographic central charge one can see that they agree exactly to leading order. As discussed in \cite{Couzens:2021veb} it is expected that higher order corrections to the gravity computation will yield an exact matching between the two quantities.

In the next subsection we summarise the results found in \cite{Faedo:2020nol}, which show that a subclass of the previous solutions with $\text{CY}_2=T^4$ can be interpreted as describing two dimensional defects within the 5d CFT dual to the AdS$_6$ background of  Brandhuber-Oz \cite{Brandhuber:1999np}.

\subsection{Defect interpretation}\label{defectAdS3}

In \cite{Faedo:2020nol} the full brane solutions whose near horizon geometries are the AdS$_3\times S^2\times T^4\times I$ backgrounds discussed in the previous subsections were constructed. They were interpreted in terms of D2-NS5-D6 branes ending on D4-D8 bound states. Furthermore, a parametrisation was obtained that allowed to relate a subclass of the AdS$_3$ geometries to 6d domain walls that asymptote locally to AdS$_6$. This allowed to propose a dual interpretation of these AdS$_3$ solutions as surface defect CFTs within the 5d Sp(N) CFT dual to the Brandhuber-Oz AdS$_6$ background. 

The brane intersection constructed in \cite{Faedo:2020nol} reads 
\begin{equation}
\label{brane_metric_D2D4NS5D6D8}
\begin{split}
d s_{10}^2&=H_{\mathrm{D}4}^{-1/2}\,H_{\mathrm{D}8}^{-1/2}\,\left[H_{\mathrm{D}6}^{-1/2}\,H_{\mathrm{D}2}^{-1/2}\,ds^2_{\mathbb{R}^{1,1}}+H_{\mathrm{D}6}^{1/2}\,H_{\mathrm{D}2}^{1/2} \,H_{\mathrm{NS}5}(dr^2+r^2d s^2_{S^2}) \right]\\
&+H_{\mathrm{D}4}^{1/2}\,H_{\mathrm{D}8}^{1/2}H_{\mathrm{D}6}^{-1/2}\,H_{\mathrm{D}2}^{-1/2} \,H_{\mathrm{NS}5}dz^2+H_{\mathrm{D}4}^{1/2}\,H_{\mathrm{D}8}^{-1/2}H_{\mathrm{D}6}^{-1/2}\,H_{\mathrm{D}2}^{1/2}(d\rho^2+\rho^2 ds^2_{\tilde S^3}) \, ,
\end{split}
\end{equation}
and
\begin{equation}
\begin{split}\label{brane_potentials_D2D4NS5D6D8}
&C_{3}=H_{\mathrm{D}8}\,H_{\mathrm{D}2}^{-1}\,\text{vol}_{\mathbb{R}^{1,1}}\wedge dz\,,\\
&C_{5}=H_{\mathrm{D}6}\,H_{\mathrm{NS}5}\,H_{\mathrm{D}4}^{-1}\,r^2\,\text{vol}_{\mathbb{R}^{1,1}}\wedge dr \wedge \text{vol}_{S^2}\,,\\
&C_{7}=H_{\mathrm{D}4}\,H_{\mathrm{D}6}^{-1}\,\rho^3\,\text{vol}_{\mathbb{R}^{1,1}}\wedge dz\wedge d\rho \wedge \text{vol}_{\tilde S^3} \,,\\
&B_{6}=H_{\mathrm{D}8}\,H_{\mathrm{D}4}\,H_{\mathrm{NS}5}^{-1}\,\rho^3\,\text{vol}_{\mathbb{R}^{1,1}}\wedge  d\rho \wedge\text{vol}_{\tilde S^3}\,,\\\vspace{0.4cm}
&e^{\Phi}=H_{\mathrm{D}8}^{-5/4}\,H_{\mathrm{D}4}^{-1/4}\,H_{\mathrm{D}6}^{-3/4}\,H_{\mathrm{NS}5}^{1/2}\,H_{\mathrm{D}2}^{1/4}\,,
\end{split}
\end{equation}
with the $C_{9}$ potential for D8 branes defining the Romans mass as $F_{0}=m$.

In this intersection the D2 and the NS5 branes are taken to be smeared over the space transverse to the D4-branes, i.e. $H_{D2}=H_{D2}(r)$ and $H_{NS5}=H_{NS5}(r)$. The Bianchi identities read
\begin{equation}\label{bianchiD2D4NS5D6D8}
 \partial_zH_{\mathrm{D}8}=m\,,\qquad H_{\mathrm{NS}5}=H_{\mathrm{D}6}=H_{\mathrm{D}2}\,,\qquad \nabla^2_{\mathbb{R}^3_r}\,H_{\mathrm{NS}5}=0\,.
\end{equation}
Imposing the relations \eqref{bianchiD2D4NS5D6D8}, the Bianchi identities for $F_{(4)}$ and the equations of motion collapse to the equation describing the D4-D8 system \cite{Imamura:2001cr},
\begin{equation}
\begin{split}\label{eomD4D8}
 &H_{\mathrm{D}8}\,\nabla_{T^4}^2\,H_{\mathrm{D}4}+\partial_{z}^2\,H_{\mathrm{D}4}=0\,.
 \end{split}
\end{equation}
Finally a particular solution can be written down as 
\begin{equation}\label{solD2D4NS5D6D8}
  H_{\mathrm{NS}5}(r)=1+\frac{Q_{\mathrm{NS}5}}{r}\,,\qquad H_{\mathrm{D}6}(r)=1+\frac{Q_{\mathrm{D}6}}{r}\,,\qquad  H_{\mathrm{D}2}(r)=1+\frac{Q_{\mathrm{D}2}}{r}\,,
 \end{equation}
where $Q_{\mathrm{D}6}=Q_{\mathrm{D}2}=Q_{\mathrm{NS}5}$ for \eqref{bianchiD2D4NS5D6D8} to be satisfied.

It was shown in \cite{Faedo:2020nol}  that this solution gives rise to the AdS$_3\times S^2\times T^4\times I$ solutions discussed in the previous subsections (restricted to the case $u'=0$) in the near horizon limit, i.e. when $r\rightarrow 0$. Furthermore, it was shown that the AdS$_3$ backgrounds asymptote locally to the AdS$_6$ vacuum associated to the D4-D8 system. This could be achieved through a change of variables that allowed to map the AdS$_3$ solutions to the uplift to massive IIA of the domain wall solutions to 6d $\mathcal{N}=(1,1)$ minimal gauged supergravity found in \cite{Dibitetto:2017klx}. These domain wall solutions were shown to asymptote locally to the AdS$_6$ vacuum of 6d supergravity, and therefore, upon uplift, to the Brandhuber-Oz AdS$_6$ solution of massive IIA supergravity.  This goes as follows.

In \cite{Dibitetto:2017klx} the following 6d background was considered,
\begin{equation}
\begin{split}\label{6dAdS3}
& ds^2_6=e^{2U(\mu)}\left(4\,ds^2_{AdS_3}+ds^2_{S^2} \right)+e^{2V(\mu)}d\mu^2\,,\\
&\mathcal{B}_{2}=b(\mu)\,\text{vol}_{S^2}\,,\\
&X_6=X_6(\mu)\,.
\end{split}
\end{equation}
This background is described by the set of BPS equations,
\begin{equation}
 \begin{split}
  U^\prime= -2\,e^{V}\,f_6\,,\qquad X_6^\prime=2\,e^{V}\,X_6^2\,D_Xf_6\,,\qquad b^\prime= \frac{e^{U+V}}{X_6^2}\,,
  \label{chargedDW6d}
 \end{split}
\end{equation}
together with the duality constraint
\begin{equation}\label{chargedDW6d1}
 b=-\frac{e^{U}\,X_6}{m}\,,
\end{equation}
and the superpotential $f_6$ 
\begin{equation}
f_6 (m,g,X_6)=\frac18 \Bigl(mX_6^{-3}+\sqrt{2}gX_6\Bigr), \qquad \text{with} \qquad m=\frac{\sqrt{2}}{3}g.
\end{equation}
This flow preserves 8 real supercharges (BPS/2 in 6d). In order to obtain an explicit solution of \eqref{chargedDW6d}, the parametrisation of the 6d geometry
\begin{equation}
 e^{-V}=2\,X_6^2\,D_Xf_6
\end{equation}
was chosen.
The system \eqref{chargedDW6d} could then be integrated out easily \cite{Dibitetto:2017klx}, to give
\begin{equation}
 \begin{split}
  e^{2U}= &\ 2^{-1/3}g^{-2/3}\,\left(\frac{\mu}{\mu^4-1}\right)^{2/3}\ , \qquad e^{2V}=8\,g^{-2}\, 
  \frac{\mu^4}{\left( \mu^4-1\right)^2}\ ,\\
   b=&\ -2^{4/3}\,3\,g^{-4/3}\,\frac{\mu^{4/3}}{(\mu^4-1)^{1/3}}\ ,\qquad \ X_6=\mu\ ,
   \label{chargedDWsol}
 \end{split}
\end{equation}
with $\mu$ running between 0 and 1.

One can see that for $\mu \rightarrow 1$ the 6d background is such that
\begin{equation}
 \begin{split}
  \mathcal{R}_{6}= -\frac{20}{3}\,g^2+ O (1-\mu)^{2/3}\,,\qquad X_6=&\ 1+ O (1-\mu)\ ,
  \label{UVchargedDW6d}
 \end{split}
\end{equation}
where $\mathcal{R}_{6}$ is the scalar curvature. These are the curvature and scalar fields reproducing the AdS$_6$ vacuum. In turn, the 2-form gauge potential gives non-zero sub-leading contributions in this limit. This implies that the asymptotic geometry for $\mu \rightarrow 1$ is only locally AdS$_6$.  In the opposite limit $\mu\rightarrow 0$, the 6d background is manifestly singular. This is due to the presence of the D2-NS5-D6 brane sources.

The uplift of the 6d domain wall solution reads
\begin{equation}
 \begin{split}\label{uplift6dDW}
  ds^2_{10}&=s^{-1/3}\,X_6^{-1/2}\,\Sigma_6^{1/2}\,e^{2U}\left(4\,ds^2_{AdS_3}+ds^2_{S^2} \right)+s^{-1/3}\,X_6^{-1/2}\,\Sigma_6^{1/2}e^{2V}d\mu^2\\
  &+2g^{-2}s^{-1/3}\Sigma_6^{1/2}\,X_6^{3/2}\,d\xi^2+2g^{-2}\,X_6^{-3/2}\,\Sigma_6^{-1/2}\,s^{-1/3}\,c^{2}\,ds^2_{\tilde S^3}\ ,\\
  F_{4}&=-\frac{4\sqrt 2}{3}\,g^{-3}\,s^{1/3}\,c^3\,\Sigma_6^{-2}\,U\,d\xi\,\wedge\,\text{vol}_{\tilde S^3}-8\sqrt{2}\,g^{-3}\,s^{4/3}\,c^4\,\Sigma_6^{-2}\,X_6^{-3}\,X_6^\prime\,d\mu\,\wedge\,\text{vol}_{\tilde S^3}\\
  &-8\,\sqrt2 \,g^{-1}\,s^{1/3}\,c\,X_6^4\,b^\prime\,e^{U-V}\,d\xi \wedge \text{vol}_{AdS_3}-8\, m\,s^{4/3}\,b\,X_6^{-2}\,e^{U+V}\,d\mu \wedge \text{vol}_{AdS_3}\,,\\
  F_{2}&=m\,s^{2/3}\,b\,\text{vol}_{S^2}\ ,\qquad H_{3}=s^{2/3}\,b^\prime\,d\mu \wedge \text{vol}_{S^2}+\frac{2}{3}\,s^{-1/3}\,c\,b\,d\xi \wedge \text{vol}_{S^2}\ ,\\
  e^{\Phi}&=s^{-5/6}\,\Sigma_6^{1/4}\,X_6^{-5/4}\ ,\qquad F_{(0)}=m\,,
 \end{split}
\end{equation}
with $c=\cos\xi$, $s=\sin \xi\,,\,\,\Sigma_6=X_6\,c^2+X_6^{-3}\,s^2$ and $U$ given by 
\begin{equation}
U=X_6^{-6}s^2-3X_6^2 c^2+4X_6^{-2}.
\end{equation}
It was shown in  \cite{Faedo:2020nol}  that the background \eqref{uplift6dDW} takes exactly the form of the AdS$_3$ solutions defined by \eqref{seedNS}-\eqref{RRsector-seed}, upon the change of coordinates
\begin{equation}\label{coord6dAdS6}
  z=\frac{3\,s^{2/3}\,e^U\,X_6}{\sqrt{2}\,g\, Q_{\mathrm{NS}5}}\,, \qquad \rho=\frac{\sqrt 2\,c\,e^{3U/2}}{g\,Q_{\mathrm{NS}5}^{3/2}\,X_6^{1/2}}\,.
\end{equation}
The AdS$_3$ solution is then specified by the functions
\begin{equation} \label{restH8H4}
 H_{\mathrm{D}8}=\frac{s^{2/3}\,e^U\,X_6}{Q_{\mathrm{NS}5}}\,,\qquad H_{\mathrm{D}4}=\frac{Q_{\mathrm{NS}5}^5\,e^{-5U}}{\Sigma_6}\,,
\end{equation}
which can be shown to satisfy the Bianchi identities given by \eqref{eqsmotion}, with $H_{D8}=h_8$ and $H_{D4}=h_4$.

We have thus shown that the AdS$_3$ backgrounds describing the near-horizon limit of D2-NS5-D6 branes ending on the D4-D8 brane system, reproduce locally the AdS$_6$ vacuum of \cite{Brandhuber:1999np} for  $H_{\mathrm{D}8}$, $H_{\mathrm{D}4}$ given by (\ref{restH8H4}). This vacuum geometry comes out thanks to a non-linear mixing of the $(z,\rho)$ coordinates, that relates the near-horizon geometry to a 6d domain wall admitting AdS$_6$ in its asymptotics. The presence of the 2-form does not allow however to globally recover the vacuum in this limit. This is seen explicitly at the level of the uplift \eqref{uplift6dDW}, where one notes that the $F_{2}$ and $H_{3}$ fluxes break  the isometries of the D4-D8 vacuum. This is the manifestation of the D2-NS5-D6 defect, that underlies as well the singular behaviour of the 6d domain wall in its IR regime.

\vspace{0.5cm}

In the next section we summarise a new class of AdS$_2$ solutions to massive Type IIA supergravity obtained from the solutions reviewed in this section via a double analytical prescription. We describe the superconformal quantum mechanics dual to these solutions, together with a very similar defect interpretation within AdS$_6$ to the one presented in this subsection.

\section{AdS$_2$/CFT$_1$ with 4 supersymmetries}\label{TypeC}

We start this section reviewing the new class of  AdS$_2\times$S$^3\times$CY$_2\times$I  solutions to massive Type IIA supergravity studied in \cite{Lozano:2020bxo,Lozano:2020sae}. These solutions were obtained via a double analytical continuation from the solutions reviewed in Section \ref{seed}. This double analytical continuation changes the AdS$_3$ and S$^2$ factors of the backgrounds in \eqref{seedNS}-\eqref{RRsector-seed} as,
\begin{equation}
\begin{split}
&\text{d}s_{\text{AdS}_3}^2\to-\text{d}s_{\text{S}^3}^2	,\qquad \text{d}s_{\text{S}^2}^2\to-\text{d}s_{\text{AdS}_2}^2.
\label{diagramaeqI}
\end{split}
\end{equation}
In order to get well-defined supergravity fields the $h_8,h_4$ and $u$ functions need to be also analytically continued as,
\begin{equation}
\begin{split}
u\to -iu,\qquad h_4\to ih_4, \qquad h_8\to ih_8, 
\label{diagramaeqII}
\end{split}
\end{equation}
together with $z\to i z$. In this way, one finds a class of
AdS$_2\times$S$^3\times$CY$_2\times$I solutions to massive Type IIA supergravity with 4 supercharges, with NS-NS sector given by\footnote{As in section 2 we have restricted to the case in which the closed and anti-self dual 2-form living on the $\text{CY}_2$ vanishes.} 
 \begin{equation}\label{NSsector-C}
 	\begin{split}
 		&\text{d}s^2 = \frac{u}{\sqrt{h_4 h_8}} \left( \frac{h_4 h_8}{\tilde{\Delta}} \text{d}s^2_{\text{AdS}_2} + \text{d}s^2_{\text{S}^3}\right) + \sqrt{\frac{h_4}{h_8}} \text{d}s^2_{\text{CY}_2} + \frac{\sqrt{h_4 h_8}}{u} \text{d} z^2 \, ,\qquad \tilde{\Delta}=4 h_4 h_8 - u'^2, 
 		\\
 		&e^{- 2\Phi}= \frac{h_8^{3/2}\tilde{\Delta}}{4 h_4^{1/2} u} \, ,\qquad\qquad 
 		B_2 = - \frac{1}{2}\bigg( z + \frac{u u'}{\tilde{\Delta}} \bigg)\text{vol}_{\text{AdS$_2$}},\, 
 	\end{split}
 \end{equation}
and RR fluxes,
\begin{equation}\label{RRsecto-C}
	\begin{split}
		F_{0} &= h_8' \, , \qquad F_{2} = - \frac{1}{2} \Big( h_8 + \frac{h_8' u' u}{\tilde{\Delta}} \Big) \text{vol}_{\text{AdS$_2$}} \, , \\
		F_{4} &= \left( - \text{d} \bigg( \frac{u'u}{2 h_4} \bigg) + 2 h_8 \text{d} z \right) \wedge \text{vol}_{\text{S}^3} - \partial_z h_4 \text{vol}_{\text{CY$_2$}}-h_8(*_4 d_4 h_4)\wedge dz  \, .
	\end{split}
\end{equation}
These backgrounds are associated to D0-F1-D4-D4$'$-D8 brane intersections that preserve $\mathcal{N}=4$ supersymmetries in one dimension. The corresponding brane set-up is depicted in Table \ref{branesetup-TypeC}.

\begin{table}[t]
 	\begin{center}
 		\begin{tabular}{|c|c|c|c|c|c|c|c|c|c|c|}
 			\hline & $x^0$ & $x^1$ & $x^2$ & $x^3$ & $x^4$ & $x^5$ & $x^6$ & $x^7$ & $x^8$ & $x^9$\\ 
 			\hline
 			D0 & x & & & & &  & & & &  \\
 			\hline
 			D4 &  x &x &x  &x  &x  & & $$ & $$ & $$ &  \\
 			\hline
 			D$4'$ & x & $$ & $$ & $$ & $$ & $$ &x &x &x &x \\
 			\hline
 			D8 & x & x & x & x & x & & x & x & x &x  \\
 			\hline F1 &x & $$ & $$ & $$ & $$ & x& $$ & & & $$ \\ \hline
 		\end{tabular}
 		\caption{Brane set-up associated to the solutions \eqref{NSsector-C}-\eqref{RRsecto-C}.  $x^0$ is the time direction of the ten dimensional spacetime, $x^1, \dots , x^4$ are the coordinates spanned by the CY$_2$, $x^5$ is the direction where the F1-strings are stretched, and $ x^6, x^7, x^8, x^9$ are the coordinates where the SO$(4)$ R-symmetry is realised.}
 		\label{branesetup-TypeC}
 	\end{center}
 \end{table}
 As in the AdS$_3\times$S$^2$ solutions we restrict to the case in which $h_4$ does not depend on the coordinates of the $\text{CY}_2$. In this case 
the functions $h_8$, $h_4$  and $u$ have support on $z$, and satisfy the constraints imposed for supersymmetry and the Bianchi identities, away from localised sources, given by expressions  \eqref{eqsmotion1} and \eqref{eqsmotion}, with $\nabla^2_{\text{CY}_2}h_4=0$. Thus $h_8$, $h_4$ and $u$ are again linear functions of $z$.

The Page fluxes are given by,
\begin{equation}\label{fluxPage-C}
	\begin{split}
		&\hat{F}_{0} = h_8' \, , \qquad\qquad \hat{F}_{2} =  -\frac{1}{2} \Big( h_8 -h_8'(z-2\pi k) \Big) \text{vol}_{\text{AdS$_2$}} \, , \\
		&\hat{F}_{4} = \left(2 h_8 \text{d} z  - \text{d} \bigg( \frac{u'u}{2 h_4} \bigg) \right) \wedge \text{vol}_{\text{S$^3$}}  - h_4' \text{vol}_{\text{CY$_2$}} \,,
	\end{split}
\end{equation}
 where we have included large gauge transformations of $B_2$ of parameter $k$, $B_2\to B_2+\pi k\text{vol}_{\text{AdS}_2}$, as discussed in \cite{Lozano:2020sae}.
 
 In the next subsection we summarise the dual SCQM of the backgrounds \eqref{NSsector-C}-\eqref{RRsecto-C} for the choice of piecewise linear functions \eqref{profileh4sp}-\eqref{profileh8sp}. We consider the case $u'=0$ and discuss a concrete example with $u'\neq 0$, constructed in \cite{Ramirez:2021tkd}, in subsection \ref{NATD}.

 
 \subsection{The dual quiver quantum mechanics}\label{CFT-typeC}

In \cite{Lozano:2020sae} a proposal for a superconformal quantum mechanics living in the D0-D4-D$4'$-D8-F1 brane set-up depicted in Table \ref{branesetup-TypeC} 
was given in terms of a generalisation of the ADHM quantum mechanics described in \cite{Kim:2016qqs}\footnote{And of the quiver proposals discussed in  \cite{Assel:2018rcw,Assel:2019iae}.}. The quantum mechanics was interpreted as describing the interactions between brane instantons and Wilson lines in the five dimensional theory with eight Poincar\'e supersymmetries living in the D4'-D8 brane intersection. For this purpose the complete D0-D4-D$4'$-D8-F1 brane system was split into two subsystems, D4-D$4'$-F1 and D0-D8-F1, that were first studied separately.
The first subsystem was interpreted as describing BPS F1 Wilson lines introduced in the 5d theory living in the D4'-branes by D4-branes \cite{Tong:2014cha}.  Similarly, the D0-D8-F1 subsystem was interpreted as describing F1 Wilson lines introduced in the worldvolume of the D8-branes by D0-branes \cite{Chang:2016iji}. Indeed,  
both subsystems are displayed exactly as in the D3-D5-F1 brane configuration that describes Wilson lines in antisymmetric representations in 4d $\mathcal{N}=4$ SYM, studied in \cite{Yamaguchi:2006tq,Gomis:2006im}. 
 \begin{figure}[t]
	\centering
	\includegraphics[scale=0.75]{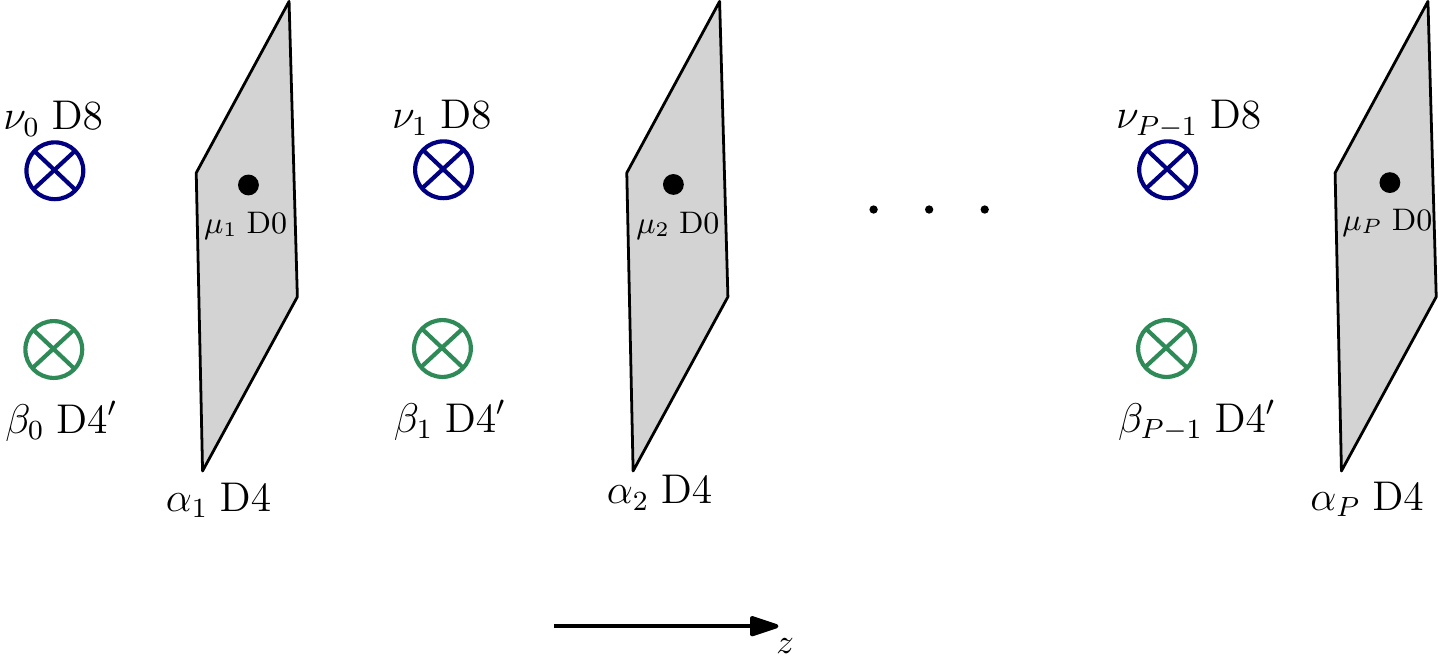}
	\caption{Hanany-Witten brane set-up associated to the solutions \eqref{NSsector-C}-\eqref{RRsecto-C} for the choice of $h_4$, $h_8$ linear functions given by \eqref{profileh4sp}-\eqref{profileh8sp}.}
	\label{HWTypeC}
\end{figure}

The quantised charges derived from the Page Fluxes \eqref{fluxPage-C} for the piecewise linear functions defined in \eqref{profileh4sp}-\eqref{profileh8sp} are given by,
\begin{eqnarray}
&&Q_{\text{D4}}^{e(k)}=\alpha_k=\sum_{j=0}^{k-1}\beta_j, \qquad Q_{\text{D0}}^{e (k)}=\mu_k=\sum_{j=0}^{k-1}\nu_j\nonumber\\
&&Q_{\text{D4}'}^{m (k)}=\beta_k,\qquad Q_{\text{D8}}^{m(k)}=\nu_k,\qquad Q_{\text{F1}}^{e(k)}=1, \label{quantised-charges}
\end{eqnarray}
in the $[2\pi k, 2\pi (k+1)]$  interval. 
Here, the superscripts $e$ and $m$ indicate electric and magnetic charges, as discussed in  \cite{Lozano:2020sae}. 

\begin{figure}[t]
	\centering
	\includegraphics[scale=0.7]{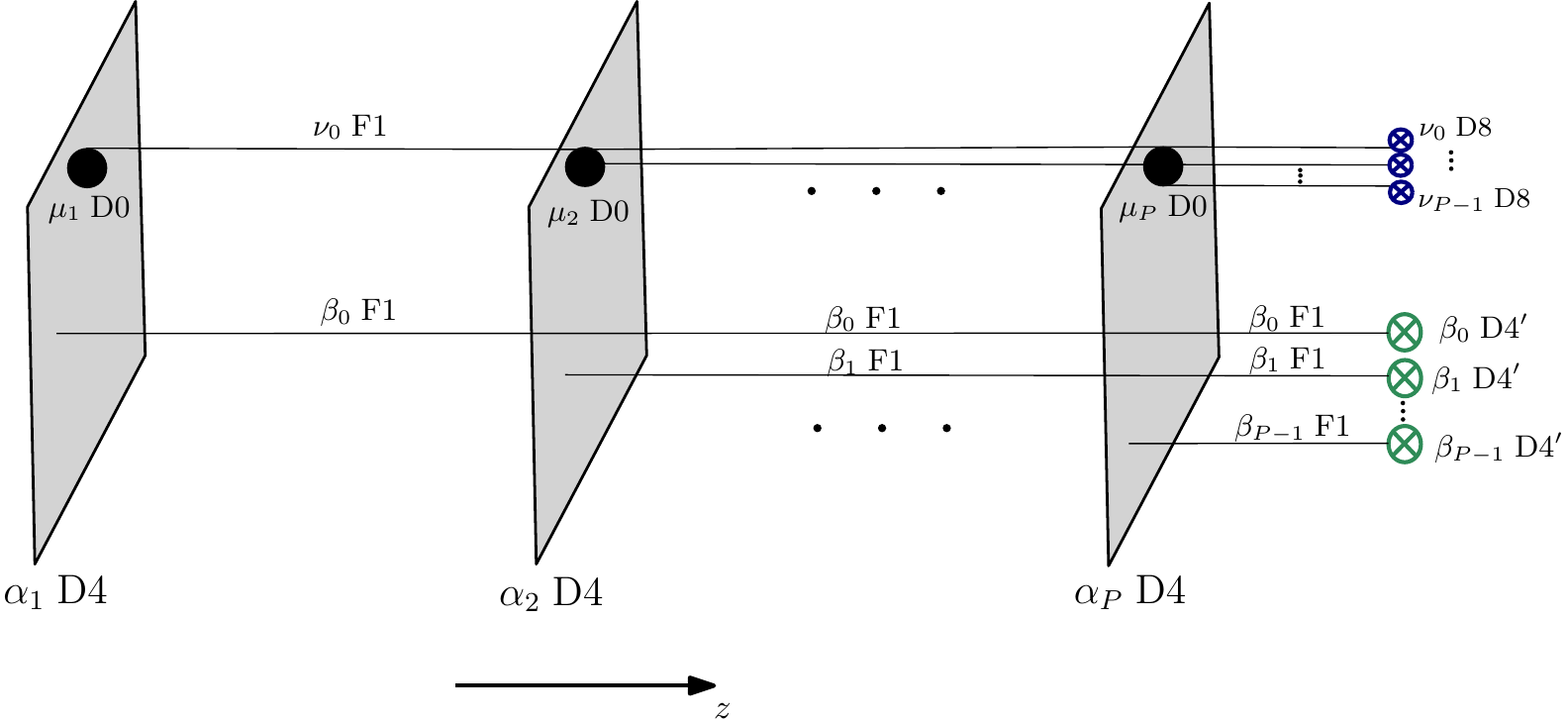}
	\caption{Brane configuration equivalent to the Hanany-Witten brane set-up depicted in Figure \ref{HWTypeC} after a T+S+T duality transformation and suitable Hanany-Witten moves.}
	\label{HWTypeCbis}
\end{figure}

The previous information can be summarised in the Hanany-Witten brane set-up depicted in Figure \ref{HWTypeC}. In order to see the interpretation as Wilson lines one can map the brane configuration onto a 
F1-D3-NS5-NS7-D1 system in Type IIB via a T+S duality transformation, perform suitable Hanany-Witten moves and then go back to Type IIA via a further T-duality. This set of operations is carefully explained in \cite{Lozano:2020sae}. One then obtains 
the configuration depicted in Figure \ref{HWTypeCbis}, which can be interpreted as describing U$(\alpha_k)$ and U$(\mu_k)$ Wilson lines in the completely antisymmetric representations $(\beta_0,\beta_1,\dots,\beta_{k-1})$ of U$(\alpha_k)$ and 
$(\nu_0,\nu_1,\dots,\nu_{k-1})$ of U$(\mu_k)$, respectively.
Given that the Wilson lines are in the completely antisymmetric representations the D4-D4'-F1 and D0-D8-F1 subsystems  describe in fact  baryon vertices \cite{Witten:1998xy}.

  This is consistent with an interpretation of the AdS$_2$ solutions as describing backreacted baryon vertices within the 5d 
 ${\cal N}=1$ QFT living in the D4'-D8 branes. In this interpretation the dual SCQM arises in the very low energy limit of a  D4'-D8 brane configuration, dual to a 5d QFT, where D4 and D0 brane baryon vertices are introduced. In the low energy limit the gauge symmetry on both the D4' and D8 branes becomes global, shifting them from colour to flavour branes, with the D4 and D0 defect branes becoming the new colour branes of the backreacted configuration.  
 This defect interpretation is in  agreement with the results found in  \cite{Faedo:2020nol}, that we summarise in subsection \ref{defectAdS2},  where the AdS$_2$ geometries were shown to asymptote locally to the AdS$_6$ background of Brandhuber-Oz \cite{Brandhuber:1999np}.
 
\begin{figure}[t]
	\centering
	\includegraphics[scale=0.55]{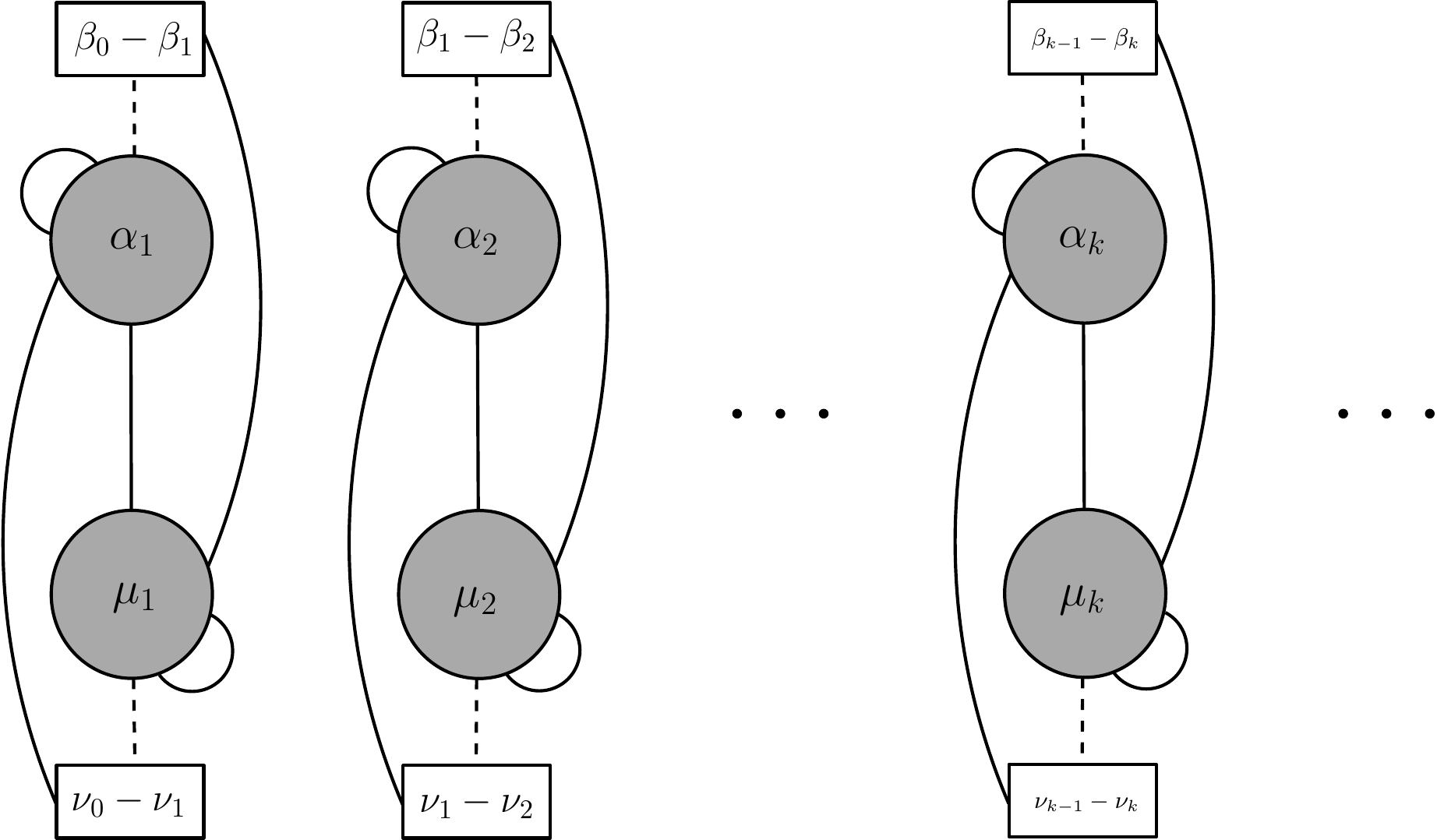}
	\caption{One dimensional quiver field theory whose IR limit is dual to the AdS$_2$ backgrounds \eqref{NSsector-C}-\eqref{RRsecto-C}.}
	\label{QuiverTypeC}
\end{figure}

The superconformal quantum mechanics dual to the AdS$_2$ solutions was analysed in detail in \cite{Lozano:2020sae}.
In the UV it is encoded in the quiver construction depicted in Figure \ref{QuiverTypeC}. In these quivers the gauge groups are associated to the colour D0- and D4-branes and  the flavour groups to the D$4'$- and D8-branes. The quantised charges are the ones computed in  (\ref{quantised-charges}). The dynamics is described in terms of (4,4) vector multiplets (circles), (4,4) hypermultiplets in the adjoint representations (semicircles) and (4,4) hypermultiplets in the bifundamental representations (vertical lines). The connection between colour and flavour branes is through twisted (4,4)  bifundamental hypermultiplets (bent lines) and (0,2) bifundamental Fermi multiplets (dashed lines). This  follows directly from the analysis in Appendix B of \cite{Lozano:2020sae}. Note that as in that reference we use 2d $\mathcal{N}=(0,4)$ notation to actually refer to the 1d $\mathcal{N}=4$ multiplets.

Checking the agreement between the field theory and holographic central charges in SCQMs is less direct than in the 2d cases discussed in the previous section. Indeed, in a one dimensional field theory the energy momentum tensor has only one component, which must thus vanish if the theory is conformal. A possible way to interpret the central charge is then as counting the ground states of the conformal quantum mechanics. This quantity is the one that is compared to the holographic central charge, which can be computed as usual from the volume of the internal manifold. In our case it reads
 \begin{equation}
 	\label{CC-TypeC}
 	c_{\text{hol}}=\frac{3 V_{int}}{4\pi G_N}=\frac{3}{\pi}\int_{0}^{2\pi(P+1)}h_4h_8\;\text{d}z.	
 \end{equation}
As discussed in \cite{Lozano:2020sae}, this result suggests that the same expression used in section \ref{CFT-seed} for the  central charge of a 2d $\mathcal{N}=(0,4)$ CFT gives the number of ground states of a $\mathcal{N}=4$ SCQM, with $n_{hyp}$ counting now the number of (untwisted) 1d $\mathcal{N}=4$ hypermultiplets and $n_{vec}$ the number of 1d $\mathcal{N}=4$ vector multiplets. Again, perfect agreement  was found in the holographic limit between this definition of the quantum mechanics central charge and the holographic central charge given by \eqref{CC-TypeC}.

As emphasised  in \cite{Lozano:2020sae} this is a striking result, since the superconformal quantum mechanics dual to the AdS$_2$ solutions does not have a priori any relation to a 2d CFT. Comparing with results in the literature for the dimension of the Higgs branch of $\mathcal{N}=4$ quantum mechanics with gauge groups $\prod_v$ U$(N_v)$ connected by bifundamentals \cite{Denef:2002ru}, one can see that  the expression $c_{cft}=6(n_{hyp}-n_{vec})$ may be interpreted as an extension of the formulas therein to more general  $\mathcal{N}=4$ quivers including flavours. This is an interesting result that deserves further investigation.

\subsection{A concrete AdS$_2$ example from non-Abelian T-duality} \label{NATD}

In this section we review a concrete  example in the previous classification constructed in \cite{Ramirez:2021tkd} via non-Abelian T-duality acting on a non-compact, freely acting, SL$(2,\mathbf{R})$ group.

The idea to study non-Abelian T-duality as a solution generating technique in supergravity was put forward in \cite{Sfetsos:2010uq}. Since then, NATD has been successfully used in the context of holography to generate new AdS backgrounds (see \cite{Sfetsos:2010uq}-\cite{Itsios:2017cew} for a set of interesting examples
and \cite{Dibitetto:2019nyz,Lozano:2021rmk} for more recent ones). Nevertheless, previous to \cite{Ramirez:2021tkd} the dualisation had been carried out with respect to a freely acting SU(2) subgroup  of the total symmetry group of the background. 

The main purpose of \cite{Ramirez:2021tkd} was to develop NATD as a solution generating technique in supergravity with respect to a freely acting non-compact SL$(2,\mathbf{R})$ group, and to apply the procedure to the D1-D5 near horizon system as an illustrative example. The  resulting geometry was shown to belong to the AdS$_2\times$S$^3\times$CY$_2$ class of solutions given by \eqref{NSsector-C}-\eqref{RRsecto-C}. This allowed to construct 
an explicit completion of the quiver quantum mechanics proposed in \cite{Lozano:2020sae}.

The starting point is a Type II background with a NS-NS sector invariant under SL$(2,\mathbf{R})\times$ SL$(2,\mathbf{R})$,
 \begin{gather}
	\label{ssm}
	\begin{split}
		ds^2&=\frac{1}{4}g_{\mu\nu}(x)L^\mu L^\nu+G_{i\mu}(x)dx^iL^\mu+G_{ij}(x)dx^idx^j,\qquad \Phi=\Phi(x)\\
		B_2&=\frac{1}{8}b_{\mu\nu}(x)L^\mu\wedge L^\nu +\frac{1}{2}B_{i\mu}(x)dx^i\wedge L^\mu+
		B_{ij}(x)dx^i\wedge dx^j,
	\end{split}
\end{gather}
where $i,j=1,2,...,7$, and $L^{\mu}$ are the SL$(2,\mathbf{R})$ left-invariant Maurer-Cartan forms given by $L^\mu=-i\textrm{Tr}(t^\mu g^{-1}\text{d}g)$. A string propagating in such background is described by a sigma model that can be  dualised with  respect to the full SL$(2,\mathbf{R})$ isometry group acting on the left (or on the right), following the rules first given in \cite{delaOssa:1992vci}. The first step is to gauge the global symmetry, replacing ordinary derivatives with covariant derivatives, $dg\to Dg=dg-Ag$. Then, the Lagrange multiplier term $ -i\textrm{Tr}(vF)$ needs to be added in order to enforce a flat connection, with   $F=dA-[A,A]$ and  $v$ a vector that takes values in the Lie algebra of the SL(2,$\mathbf{R}$) group.
After integrating by parts the Lagrange multiplier term and fixing the gauge, that we do by
setting $g=\mathbb{I}$, one obtains the NATD sigma model. As the variables parametrising the 
SL(2,$\mathbf{R}$) group are replaced by the Lagrange multipliers $\nu_i$, which by construction span the vector space $\mathbf{R}^3$, the AdS$_3$ space is replaced  in a suitable parametrisation by AdS$_2\times \mathbf{R}^+$.  More details on this dualisation can be found in  \cite{Alvarez:1993qi}.

In \cite{Ramirez:2021tkd} the dualisation was performed on the AdS$_3\times$S$^3\times$CY$_2$ geometry that arises as the near horizon limit of the D1-D5 system, 
\begin{gather}
\begin{split}
\label{AdS3S3T4}
&ds^2=4L^2 ds^2_{\text{AdS}_3}+M^2 ds^2_{\text{CY}_2}+4L^2 ds^2_{\text{S}^3},\qquad
e^{2\Phi}=1,\qquad
F_3=8L^2(\text{vol}_{\text{S}^3}+\text{vol}_{\text{AdS}_3}).
\end{split}
\end{gather}
In this case due care must be taken of the RR sector, which was dualised following the prescription in \cite{Sfetsos:2010uq} (the reader is referred to  \cite{Ramirez:2021tkd} for more details).
Parametrising the dual coordinates as $(z, \text{AdS}_2)$ the background generated reads,
\begin{gather}
\begin{split}
\label{NATD1}
ds^2&= \frac{L^2z^2}{z^2-4L^4}ds^2_{\text{AdS}_2}+4L^2ds^2_{\text{S}^3}+M^2 ds^2_{\text{CY}_2}+\frac{dz^2}{4L^2}, \qquad e^{2\Phi}=\frac{4}{L^2(z^2-4L^4)},\\
B_2&=-\frac{z^3}{2(z^2-4L^4)}\text{vol}_{\text{AdS}_2},\qquad
F_0=L^2,\qquad
F_2=-\frac{L^2z^3}{2(z^2-4L^4)}\text{vol}_{\text{AdS}_2},\\
F_4&=-L^2 (M^4 \text{vol}_{\text{CY}_2}-2 z\text{d}z\wedge\text{vol}_{\text{S}^3}).
\end{split}
\end{gather}
Notice that from the original SO$(2,2)\cong\textrm{SL}(2,\mathbf{R})_L\times\textrm{SL}(2,\mathbf{R})_R$ isometry group just one SL(2,$\mathbf{R}$) subgroup survives after the dualisation. This group is geometrically realised by a warped AdS$_2\times\mathbf{R}^+$ subspace. As anticipated, 
it is easy to see that the background \eqref{NATD1} fits locally in the class of AdS$_2$
solutions given by \eqref{NSsector-C}-\eqref{RRsecto-C}, with the choices, 
\begin{gather}
\begin{split}
\label{uNATD}
u= 4L^4 M^2 z,\qquad
h_4= L^2 M^4 z, \qquad
h_8=F_0 z\, .
\end{split}
\end{gather}
In this case due to the $z$ dependence of $u$ there is a singularity at $z=z_0=2L^2$, where the metric and dilaton behave as
\begin{equation}
\label{F1-singularity}
ds^2\sim \frac{a_1}{z-z_0}ds^2_{\text{AdS}_2}+a_2	ds^2_{\text{S}^3}+M^2ds^2_{\text{CY}_2}+a_3 dz^2,\qquad\qquad e^{\Phi}\sim a_4(z-z_0)^{-1/2}.
\end{equation}
This behaviour can be interpreted in terms of F1-strings with AdS$_2$ worldvolume smeared over the $S^3$\footnote{Note that it is also compatible with an orientifold fixed plane with F1-charge smeared on the $S^3$. The string theory interpretation of such object is however unclear.}. The metric has the correct signature and the dilaton is well-defined when $z\in [z_0,\infty)$.

The brane intersection associated to the new solution can be read from the Page fluxes, which are given by
\begin{equation}
\begin{split}
\label{Page}
\hat{F}_0=&L^2,\quad\quad\quad
\hat{F}_2=-L^2k\pi\text{vol}_{\text{AdS}_2},\quad\quad\quad\quad
\hat{F}_4=-L^2 (M^4 \text{vol}_{\text{CY}_2}-2z\;\text{d}z\wedge\text{vol}_{\text{S}^3}),
\end{split}
\end{equation}
where we have taken into account the large gauge transformations $B_2\to B_2+\pi k \text{vol}_{\text{AdS}_2}$ as in \cite{Lozano:2020sae}. These give rise to the  D0-D4-D$4'$-D8-F1 brane intersection depicted in Table \ref{branesetup-TypeC}.
The D0 and D4-branes can then be interpreted as instantons carrying electric charge,
\begin{equation}
\label{chargesNATD}
\begin{split}
	 Q_{\text{D}0}^e
	 =-k\;Q_{\text{D}8}^m,\qquad 
	Q_{\text{D}4}^e
	=-k\; Q_{\text{D}4'}^m\;,
\end{split}	
\end{equation}
while the D$4'$ and D8-branes find an interpretation as magnetically charged branes where the instantons lie, with charges
\begin{equation}
\label{chargesNATD1}
\begin{split}
	Q_{\text{D}8}^m
	=2\pi L^2,\qquad
	Q_{\text{D}4'}^m&
	=-2\pi L^2M^4,\qquad
\end{split}	
\end{equation}
 in the interval $[z_k,z_{k+1}]$. On top of this there are fundamental strings electrically charged with respect to the 3-form $H_3$,
 \begin{gather}
Q_{\text{F1}}^e=\frac{1}{(2\pi)^2}\int_{\text{AdS}_2\times\text{I}_z}H_3=\left.\frac{1}{\pi}B_2\;\right|_{z_k}^{z_{k+1}}=1.
\end{gather}
 
\begin{figure}[t]
	\centering
	\includegraphics[scale=0.7]{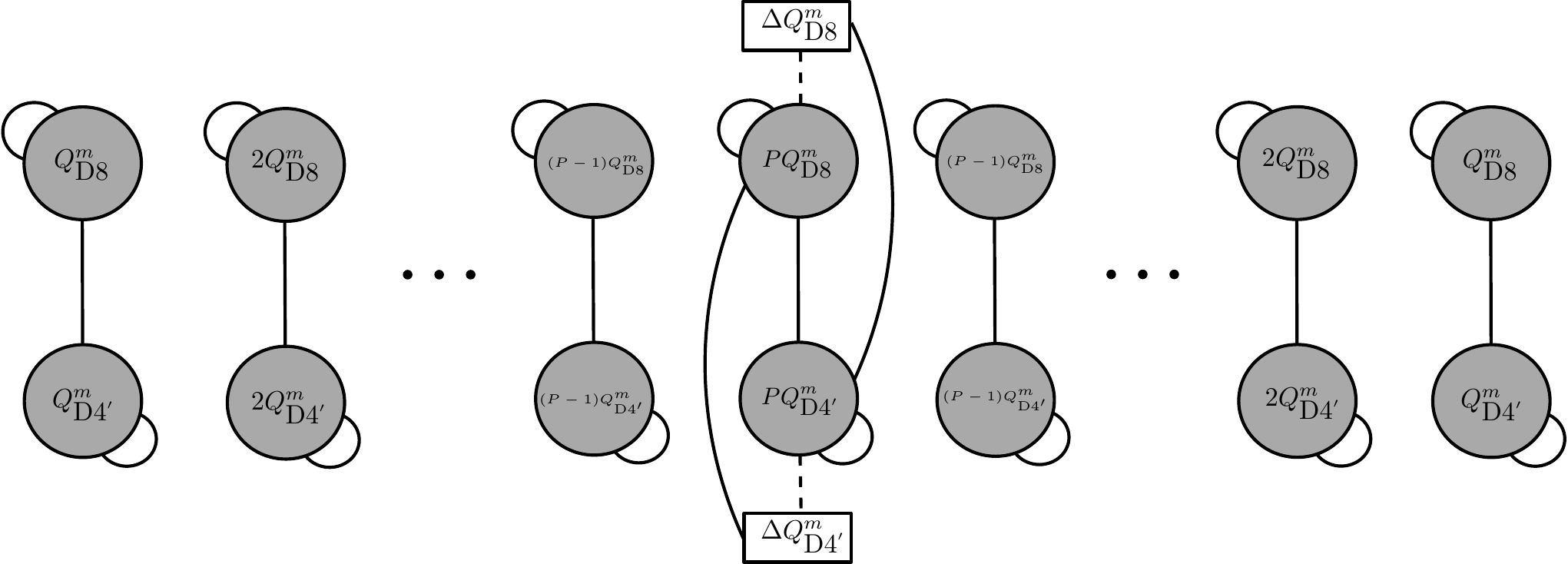}
	\caption{Completed quiver associated to the NATD solution.}
	\label{QuiverNATD}
\end{figure}

\begin{figure}[t]
	\centering
\includegraphics[width=15cm, height=8cm]{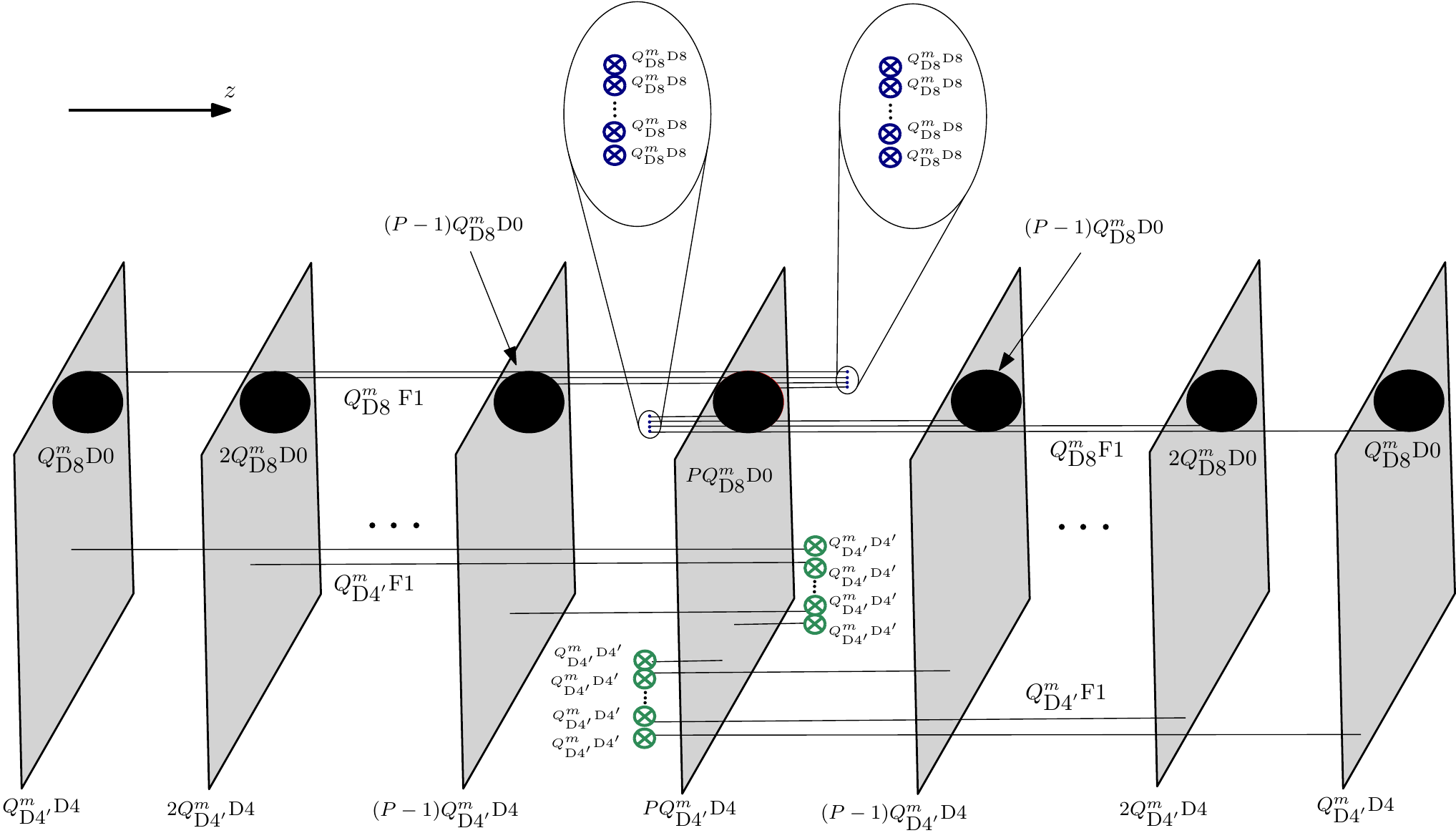}
	\caption{Brane set-up describing the baryon vertex interpretation of the quiver depicted in Figure \ref{QuiverNATD}.
	}
	\label{HW-like-NATDII} 
\end{figure}

 The holographic central charge is obtained from the volume of the internal manifold, giving
 \begin{gather}
\begin{split}
\label{hcc-NATD1}
&c_\text{hol}
=	\frac{3L^4M^4}{\pi}\int_{z_0}^\infty
(z^2-4L^4)\;\text{d}z.
\end{split}
\end{gather}
However, in order to obtain a finite value from this expression the dual background has first to be defined globally. The global completion proposed in \cite{Ramirez:2021tkd} ended the geometry with F1-strings at a certain value
 $z_{2P}$, with $P\in \mathbb{Z}$, and glued the $h_4$, $h_8$ linear functions associated to the solution at $z_P$ in a symmetric fashion
\begin{gather} \label{profileh4NATD}
\begin{split}
\frac{h_4(z)}{L^2M^4}\!
&=\!\!
                    \;\!\!\left\{ \begin{array}{cccrcl}
                      z
                        & z_0\leq z\leq z_P\;, &\\
                     z_0- (z-z_{2P})&~~ z_P\leq z \leq z_{2P},&
                                             \end{array}
\right.\quad
\frac{h_8(z)}{L^2}
=\!\!
                    \;\!\!\left\{ \begin{array}{cccrcl}
                      z
                        & z_0\leq z\leq {z_P}, &\\
                      {z_0}-  (z-z_{2P})&~~ {z_P}\leq z \leq z_{2P}.&
                                             \end{array}
\right.
\end{split}
\end{gather} 
Indeed, one can check that the NS sector is continuous at $z_P$ when $z_P=\frac{z_0+z_{2P}}{2}$, thus leading to a symmetric configuration. The quantised charges associated to this choice of linear functions, displayed in Table \ref{charges}, are such that the D0 and D4 charges increase linearly in the $0\leq k \leq P$ region while they decrease in the  $P\leq k \leq 2P$ region. 
\begin{table}[h]
\begin{center}
\begin{tabular}{|c|c|c|c|c|}
\hline
 &$Q^m_\text{D8}$ 
 &$Q^e_\text{D0}$
 &$Q^m_{\text{D4}'}$
 &$Q^e_{\text{D4}}$ \\
\hline\hline
$0\leq k \leq P$& $2\pi L^2$ &$k\;Q^m_\text{D8}$&$2\pi L^2M^4$&$k\; Q^m_{\text{D4}'}$\\
\hline
$P\leq k \leq 2P$&$2\pi L^2$ &$(2P-k)\; Q^m_\text{D8}$ &$2\pi L^2M^4$  & $(2P-k)\; Q^m_{\text{D4}'}$\\
\hline
\end{tabular}
\end{center}
\caption{Page charges of the completed non-Abelian T-dual solution --here we are expressing the absolute value of the charges}
\label{charges}
\end{table}
Given the continuity of the $h_4$ and $h_8$ functions in the two regions there are no D4'-D8 flavour branes at any of the associated nodes. The exception is at $z=z_P$, where they  jump as
\begin{equation}
\Delta Q_{\text{D$4'$}}^m=4\pi L^2 M^4, \qquad \Delta Q_{\text{D8}}^m=4\pi L^2.
\end{equation}
The associated quiver has been depicted in Figure \ref{QuiverNATD}.
The interpretation of the quiver quantum mechanics is as describing backreacted D0-D4 baryon vertices  in the completely  antisymmetric representation of the gauge groups U$(kQ_{\textrm{D8}}^m)\times$U$(kQ_{\textrm{D4}'}^m)$ associated to a 5d intersection of D4'-D8 branes. The brane set-up associated to the quiver becomes after T+S duality, suitable Hanany-Witten moves and a further T-duality, the one depicted in Figure  \ref{HW-like-NATDII}. 

One can check that, as expected, the holographic and field theory central charges, given by
\begin{eqnarray}
	c_{hol}&=&Q_{\text{D4}'}^m Q_{\text{D8}}^m(4P^3-12P+8),\\
	c_{ft}&=&  Q_{\text{D4}'}^mQ_{\text{D8}}^m (4P^3+2P),
\end{eqnarray}
coincide in the $P\rightarrow\infty$, holographic, limit.

Before we close this subsection we would like to report on recent progress in trying to connect the solution just discussed with the three dimensional black hole constructed in \cite{Alvarez:1993qi}. The aforementioned black hole was constructed by dualising the Principal Chiral Model with group manifold SL$(2,\mathbf{R})$ with respect to its whole isometry group, acting on the left. This is an AdS$_3$ geometry consisting on just metric which is not a good string theory background, since it does not satisfy the 10d equations of motion.  Given that the same holds after dualisation, the black hole geometry constructed in  \cite{Alvarez:1993qi} had limited applicability. In the remainder of this subsection we exploit the similarities between the construction carried out in  \cite{Alvarez:1993qi} and the one pursued in this subsection to try to embed the black hole geometry of \cite{Alvarez:1993qi}  in a valid string theory background. 
As we show this is not possible due to the sick behaviour of the dilaton.

Following \cite{Alvarez:1993qi} we can try to find a black hole geometry embedded in the non-Abelian T-dual solution given by \eqref{NATD1} by writing it in terms of the Lagrange multipliers, $v_i$, and defining two different parametrisations for the regions $\nu^2>0$ and $\nu^2<0$, where $v^2=\eta^{ij}v_iv_j$ with $\eta^{ij}=(+,-,+)$. These two regions are interpreted as the interior and the exterior of the black hole construction in \cite{Alvarez:1993qi}. The solution \eqref{NATD1} reads in terms of the Lagrange multipliers\footnote{In this subsection we take $L=1$.},
\begin{gather}
\begin{split}
\label{AdS3S2T4Imul}
&ds^2=\frac{(dv_1^2-dv_2^2+dv_3^2)-(v_1dv_1-v_2dv_2+v_3dv_3)^2}{1-(v_1^2-v_2^2+v_3^2)}+4 ds^2_{\text{S}^3}+M^2 ds^2_{\text{CY}_2},\\
&e^{-2\Phi}=(v_1^2-v_2^2+v_3^2)-1,\qquad B_2=\frac{v_3dv_1\wedge dv_2-v_2dv_1\wedge dv_3+v_1dv_2\wedge dv_3}{(v_1^2-v_2^2+v_3^2)-1},\\
&F_0=1,\qquad F_2=\frac{(v_3dv_1\wedge dv_2-v_2dv_1\wedge dv_3+v_1dv_2\wedge dv_3)}{(v_1^2-v_2^2+v_3^2)-1},\\
&F_4=-M^4\text{vol}_{\text{CY}_2}+8(v_1dv_1-v_2dv_2+v_3dv_3)\wedge\text{vol}_{\text{S}^3}\,.\qquad\qquad
\end{split}
\end{gather}
Following \cite{Alvarez:1993qi} we then use the parametrisations\footnote{Note that $\rho=4^{-1}L^{-2}z^2$ with $z$ in  \eqref{NATD1}.}
\begin{equation}
\begin{split}
&v^{I}_i=(\sqrt{\rho}\cos\phi\cosh\xi,\sqrt{\rho}\sinh\xi,\sqrt{\rho}\sin\phi\cosh\xi),\\ &v^{II}_i=(\sqrt{\rho}\cos\phi\sinh\xi,\sqrt{\rho}\cosh\xi,\sqrt{\rho}\sin\phi\sinh\xi)	,
\label{eq:parametrisations}
\end{split}	
\end{equation}
in the two different regions 
\begin{eqnarray}
&(v^{I})^2=\rho>0,\qquad \qquad (v^{II})^2=-\rho<0.
\end{eqnarray}
 Note that both parametrisations in \eqref{eq:parametrisations} are related under $\xi\to\xi-i\frac{\pi}{2}$ and $\rho\to -\rho$.

The geometry in region I reads,
\begin{gather}
\begin{split}
\label{AdS3S2T4IregI}
&ds_{I}^2=\frac{\rho}{1-\rho}(-d\xi^2+\cosh^2\xi \;d\phi^2)+\frac{d\rho^2}{4\rho}+M^2 ds^2_{\text{CY}_2}+4 ds^2_{\text{S}^3},\qquad e^{-2\Phi}=\rho-1,\\
&B_2=-\frac{\rho^{3/2}}{(\rho-1)}\cosh\xi\; d\phi\wedge d\xi,\qquad F_0=1,\qquad
F_2=-\frac{\rho^{3/2}}{(\rho-1)}\cosh\xi\; d\phi\wedge d\xi,\\
&F_4=-M^4\text{vol}_{\text{CY}_2}+4\; d\rho\wedge\text{vol}_{\text{S}^3},\qquad\qquad
\end{split}
\end{gather}
while in region II it is given by
\begin{gather}
\begin{split}
\label{AdS3S2T4IregII}
&ds_{II}^2=\frac{\rho}{\rho+1}(\sinh^2\xi \;d\tau^2+d\xi^2)-\frac{d\rho^2}{4\rho}+M^2 ds^2_{\text{CY}_2}+4L^2 ds^2_{\text{S}^3}\;,\qquad e^{-2\Phi}=-(\rho+1)\;,\\
&B_2=-\frac{\rho^{3/2}}{\rho+1}\sinh\xi \;d\tau\wedge d\xi,\qquad F_0=1,\qquad F_2=-\frac{\rho^{3/2}}{\rho+1}\sinh\xi \;d\tau\wedge d\xi,\\
&F_4=-M^4\text{vol}_{\text{CY}_2}-4\; d\rho\wedge\text{vol}_{\text{S}^3}.
\end{split}
\end{gather}
In turn, the Ricci scalars in both regions read,
\begin{eqnarray}
R_{I}=-\frac{\rho^2-6\rho+33}{2(\rho-1)^2},\qquad\qquad	R_{II}=-\frac{\rho^2+6\rho+33}{2(\rho+1)^2}.
\end{eqnarray}
As for the black hole in \cite{Alvarez:1993qi}  there is a singularity in region I at $\rho=1$, while the behaviour at $\rho=0$ is that of an event-horizon. This is reflected in the change of signature of the metric, going from  (+,+,-,+,+,+,+,+,+,+) in region II to  (+,-,+,+,+,+,+,+,+,+) in region I, and  (-,+,+,+,+,+,+,+,+,+) beyond $\rho=1$. A detailed study of the causal structure associated to this geometry could now be carried out following \cite{Alvarez:1993qi}. Note however that there is a fundamental obstruction that invalidates a similar analysis to that in \cite{Alvarez:1993qi}, since the dilaton is ill-defined both in region II (the would-be exterior of the black hole) and in region I when $\rho<1$ (the would-be interior of the black hole). This implies that the black hole constructed in \cite{Alvarez:1993qi}  cannot be embedded within our supergravity background. The same conclusion is reached if one attempts to embed the black hole onto a more general solution in the class reviewed in section 2.  In this case the Ricci scalar is singular when ${\tilde \Delta}$ vanishes, say at $\rho_*$. According to the interpretation in \cite{Alvarez:1993qi} $\rho\in [0,\rho_*]$ would parametrise the black hole interior, and $\rho<0$ the exterior, which is again where $\tilde \Delta<0$ and the dilaton is ill-defined. 

\subsection{Defect interpretation} \label{defectAdS2}

In this subsection we show that it is possible to provide a defect interpretation to the solutions described in this section in complete analogy with the analysis performed in subsection \ref{defectAdS3}.  In this case the brane solutions whose near horizon geometries are the AdS$_2\times S^3\times T^4\times I$ backgrounds were worked out in \cite{Dibitetto:2018gtk}, and further analysed in \cite{Faedo:2020nol}, where they were interpreted in terms of D0-F1-D4 branes ending on D4'-D8 bound states. As in subsection \ref{defectAdS3} a parametrisation was obtained that allowed to relate a subclass of the AdS$_2$ geometries to a 6d domain wall solution to 6d $\mathcal{N}= (1,1)$ minimal gauged supergravity  that asymptotes locally to AdS$_6$. This allowed to propose a dual interpretation of these AdS$_2$ solutions as line defect CFTs within the 5d Sp(N) CFT dual to the Brandhuber-Oz AdS$_6$ background. 

As in the calculation in subsection \ref{defectAdS3}, allowing the D4'-branes to be completely localised in their transverse space it is possible to recover a near-horizon geometry describing a D4'-D8 system wrapping an AdS$_2\times S^3$ geometry, to which D0-F1-D4 branes need to be added to preserve supersymmetry \cite{Dibitetto:2018gtk}. The near-horizon reads
\begin{equation}\label{D8D4D0F1D4'-nh}
ds_{10}^{2}  =  H_{\textrm{D}4'}^{-1/2}H_{\textrm{D}8}^{-1/2}\left[Q_{1} \left(ds_{\textrm{AdS}_{2}}^{2}+4ds_{S^{3}}^{2}\right)+H_{\textrm{D}4'}H_{\textrm{D}8}dz^{2}+
H_{\textrm{D}4'}\,\left (d\rho^{2}+\rho^{2}\,ds_{\tilde{S}^{3}}^{2}\right)\right] \, ,
\end{equation}
with $Q_1$ a parameter related to the defect charges of D0-F1-D4 branes.
One can check that this background is included in the classification reviewed in this section, for $\text{CY}_2=T^4$ locally and $u^\prime=0$.

As already mentioned, the previous brane intersection was linked  to a 6d charged domain wall characterised by an AdS$_2$ slicing flowing asymptotically to the  AdS$_6$ vacuum of 6d Romans supergravity. This domain wall is of the form
\begin{equation}
\begin{split}\label{6dAdS2}
& ds^2_6=e^{2U(\mu)}\left(ds^2_{AdS_2}+4ds^2_{S^3} \right)+e^{2V(\mu)}d\mu^2\,,\\
&B_{2}=b(\mu)\,\text{vol}_{AdS_2}\,,\\
&X_6=X_6(\mu)\,,
\end{split}
\end{equation}
and, consistently with the whole picture, can be obtained through double analytical continuation from the domain wall solution in \eqref{6dAdS3}. The BPS equations for this background preserve 8 real supercharges and take the same form of \eqref{chargedDW6d} and \eqref{chargedDW6d1}. In analogy with the AdS$_3$ analysis, the 6d solution \eqref{6dAdS2} reproduces locally in the limit $\mu \rightarrow 1$ the geometry of the AdS$_6$ vacuum, together with a singularity in the $\mu \rightarrow 0$ limit. Using the uplift formulas  to massive IIA given in  \cite{Faedo:2020nol} one can check that the resulting domain wall solution in 10d is related to the near horizon geometry \eqref{D8D4D0F1D4'-nh} through the change of coordinates \cite{Dibitetto:2018gtk}
\begin{equation}
  z=\frac{3\,s^{2/3}\,e^U\,X_6}{\sqrt{2}\,g\, Q_1^{1/2}}\,, \qquad \rho=\frac{\sqrt 2\,c\,e^{3U/2}}{g\,Q_1^{3/4}\,X_6^{1/2}}\,.
\label{coordchangeAdS2}
\end{equation}
The AdS$_2$ solution is then specified by
\begin{equation} \label{H8H4AdS2}
 H_{\mathrm{D}8}=\frac{s^{2/3}\,e^U\,X_6}{Q_1^{1/2}}\,,\qquad H_{\mathrm{D}4'}=\frac{Q_{1}^{5/2}\,e^{-5U}}{\Sigma_6}\,,
\end{equation}
with $h_8=H_{\mathrm{D}8}$ and $h_4=H_{\mathrm{D}4'}$.
These conditions are analogous to~\eqref{coord6dAdS6}-\eqref{restH8H4} for AdS$_3$, which is obviously related  to the fact that the AdS$_2$ solutions and the AdS$_3$ backgrounds are related by double analytical continuation.
In this case the solution is interpreted as a D0-F1-D4 line defect within the 5d Sp(N) fixed point theory.

\section{Discussion}\label{discu}

In these proceedings we have reviewed recent progress in the construction of AdS$_3$/CFT$_2$ and 
AdS$_2$/CFT$_1$ dual pairs in massive Type IIA string theory. These dual pairs represent new well-controlled string theory settings where the microscopical counting program of five and four dimensional black holes can be further developed. On a different note, our solutions allow for a defect interpretation in terms of surface or line defects within the 5d Sp(N) fixed point theory. In general grounds having at our disposal the holographic description of these defect CFTs allows to apply holographic methods to the computation of central charges, correlators and other observables of the defect CFT.

Notably, other AdS/CFT pairs have been constructed in the recent literature that can also be taken as set-ups where to carry out the microscopical counting program of black holes as well as the holographic study of defect CFTs. The most direct extensions of the solutions here presented are the $\text{AdS}_3\times S^2\times M_4\times I$ and $\text{AdS}_2\times S^3\times M_4\times I$ solutions to massive IIA supergravity with $M_4$ a K\"ahler manifold, constructed in \cite{Lozano:2019emq,Lozano:2020bxo,Lozano:2020sae}. These solutions have been left out of our analysis because the corresponding field theory duals have only been partially explored or not explored at all. In the $\text{AdS}_3$ case, when $I=S^1$ and there are no D4-branes present, these solutions are related to the class discussed in \cite{Couzens:2017way} via Abelian T-duality. Therefore, the dual CFTs are described in terms of D3-branes wrapping complex curves in elliptically fibrered $\text{CY}_3$ manifolds. More general field theory settings related to these solutions have not yet been explored, neither have their possible realisations as defects within higher dimensional CFTs. These constitute interesting new research avenues to explore.

In \cite{Lozano:2020bxo} $\text{AdS}_3\times S^3\times M_4\times I$ solutions to M-theory where $M_4$ is either a $\text{CY}_2$ or a K\"ahler manifold have been constructed. These solutions preserve the same number of supersymmetries as the solutions here presented, and for $M_4=\text{CY}_2$ are dual to quiver CFTs similar to the ones reviewed in subsection 2.1, which in this case describe M-strings (see \cite{Haghighat:2013tka,Gadde:2015tra}). In \cite{Faedo:2020nol} it was shown that a subset of these solutions can be interpreted as surface defects within 6d (1,0) CFTs living in M5-branes probing ALE singularities. Moreover, upon reduction these solutions give rise to a new class of   $\text{AdS}_3\times S^3\times S^2\times \Sigma_2$ solutions, with $\Sigma_2$ a 2d Riemann surface, that can be interpreted as defects within the 6d (1,0) CFT dual to the AdS$_7$ solution to massless Type IIA supergravity. The description of the dual 2d CFT in terms of quivers embedded in the 6d quiver associated to the D6-NS5 intersection was also worked out in \cite{Faedo:2020nol}. Similarly, new classes of $\text{AdS}_3\times S^2\times S^2\times S^1\times \Sigma_2$ solutions with the same number of supersymmetries have been constructed in Type IIB supergravity \cite{Faedo:2020lyw}, some of which admit a defect interpretation within the 5d Sp(N) fixed point theory (this time realised in a Type IIB brane intersection) and/or describe holographic duals of D3-brane boxes, as the ones discussed in \cite{Hanany:2018hlz}. The readers are referred to \cite{Faedo:2020lyw} for the details of these constructions. More recently, the parallel in Type IIB of the general classification of AdS$_3$ spaces with $\mathcal{N}=(0,4)$ supersymmetries in \cite{Lozano:2019emq} has been carried out in \cite{Macpherson:2022sbs}. The readers can again find the details of these constructions in the original reference. An interesting open line to explore is the construction of the 2d CFTs dual to these solutions, along the lines of \cite{Lozano:2019jza,Lozano:2019zvg,Couzens:2021veb}.

Similarly, new classes of AdS$_2$ solutions in Type IIB with 4 supersymmetries have been constructed in \cite{Lozano:2020txg}, acting with Abelian T-duality on the AdS$_3$ subspace of the solutions reviewed in section 2. These solutions are dual, by construction \cite{Balasubramanian:2003kq,Balasubramanian:2009bg}, to SCQMs realised as discrete light-cone compactifications of the 2d dual CFTs reviewed in this paper. Further solutions of the type $\text{AdS}_2\times S^2\times \text{CY}_2 \times \Sigma_2$ with $\Sigma_2$ an annulus have been constructed in \cite{Lozano:2021rmk}, via Abelian T-duality acting on the AdS$_2$ solutions reviewed in section 3. The SCQMs dual to these solutions are thus the same as the ones reviewed in that section, and allow for a similar defect interpretation, this time as backreacted D1-D3 baryon vertices within the 5d Sp(N) fixed point theory, now realised on a D5-NS5-D7 brane web. More recently, new classes of AdS$_2$ solutions with the same number of supersymmetries have been constructed in both Type IIA and Type IIB supergravities that allow a description in terms of backreacted baryon vertices within 4d $\mathcal{N}=4$ SYM or orbifolds thereof. In this case the solutions are asymptotically locally $\text{AdS}_5\times S^5/\mathbb{Z}_k$ (or its Abelian T-dual in the IIA case). The reader is referred to \cite{Lozano:2021fkk} for more details on these constructions.

An obviously interesting avenue to pursue is to investigate the CFT duals to the broader class of AdS$_3$ solutions constructed in \cite{Lozano:2019emq} for which there is a dependence on the internal structure of the $\text{CY}_2$ manifold. This would allow to extend the 2d and 1d CFTs discussed in this paper by further exploiting the interplay between string theory dualities and the AdS/CFT correspondence, as described in the previous paragraphs. We expect to report progress in these directions in the near future.

\subsection*{Acknowledgements}

We would like to thank Chris Couzens, Federico Faedo, Niall Macpherson, Carlos Nunez, Stefano Speziali and Stefan Vandoren for collaboration in some of the results reviewed in these proceedings. YL and AR are partially supported by the Spanish government grant PGC2018-096894-B-100. AR is partially supported by the Heising-Simons Foundation, the Simons Foundation, and the National Science Foundation Grant No. NSF PHY-1748958. The work of NP is supported by the Israel Science Foundation (grant No. 741/20) and by the German Research Foundation through a German-Israeli Project Cooperation (DIP) grant "Holography and the Swampland".

\end{document}